\documentclass[fleqn,usenatbib]{mnras}

\usepackage{newtxtext,newtxmath}
\usepackage[T1]{fontenc}

\DeclareRobustCommand{\VAN}[3]{#2}
\let\VANthebibliography\thebibliography
\def\thebibliography{\DeclareRobustCommand{\VAN}[3]{##3}\VANthebibliography}

\usepackage[final]{graphicx}	% Including figure files
\usepackage{amsmath}	% Advanced maths commands
\usepackage{subcaption} % For the figures
\usepackage{siunitx}
\usepackage{float}

\title{Comparison of distance measurements to dust clouds using GRB X-ray halos and 3D dust extinction}

\author[B. \v Siljeg et al.]{
B. \v Siljeg$^{1,2,3}$, \v Z. Bo\v snjak$^{4}$\thanks{E-mail: zeljka.bosnjak@fer.hr}, V. Jeli\' c$^{5}$\thanks{E-mail: vibor@irb.hr}, A. Tiengo$^{6,7}$, F. Pintore$^{8}$ and A. Bracco$^{9}$
\\ \\
$^{1}$Faculty of Science, University of Zagreb, 10000 Zagreb, Croatia\\
$^{2}$ASTRON, PO Box 2, 7990 AA, Dwingeloo, The Netherlands\\
$^{3}$ Kapteyn Astronomical Institute, University of Groningen, P.O. Box 800, 9700 AV, Groningen, The Netherlands\\
$^{4}$Faculty of Electrical Engineering and Computing, University of Zagreb, 10000 Zagreb, Croatia\\
$^{5}$Rudjer Bo\v skovi\' c Institute, Bijeni\v cka cesta 54, 10000 Zagreb, Croatia\\
$^{6}$INAF – IASF Milano, Via Corti 12, I-20133 Milano, Italy\\
$^{7}$Scuola Universitaria Superiore IUSS Pavia, Piazza della Vittoria 15, I-27100 Pavia, Italy\\
 $^{8}$Istituto Nazionale di Astrofisica, Istituto di Astrofisica Spaziale e Fisica Cosmica di Palermo, via U. La Malfa 153, 90146 Palermo, Italy \\
$^{9}$Laboratoire de Physique de l'Ecole Normale Sup\'erieure, ENS, Universit\'e PSL, CNRS, Sorbonne Universit\'e, Universit\'e de Paris, F-75005 Paris, France\\
}

\date{Accepted 2023 September 21. Received 2023 September 21; in original form 2023 July 17.}

\pubyear{2023}

\begin{document}
\label{firstpage}
\pagerange{\pageref{firstpage}--\pageref{lastpage}}
\maketitle

% Abstract of the paper
\begin{abstract}
X-ray photons from energetic sources such as gamma-ray bursts (GRBs) can be  scattered on dust clouds in the Milky Way, creating a time-evolving halo around the GRB position. X-ray observations of such halos allow the measurement of dust clouds distances in the Galaxy on which the scattering occurs. We present the first systematic comparison of the distances to scattering regions derived from GRB halos with the 3D dust distribution derived from recently published optical-to-near infrared extinction maps. GRB halos were observed around 7 sources by the {\it Swift} XRT and the {\it XMM-Newton} EPIC instruments, namely GRB 031203, GRB 050713A, GRB 050724, GRB 061019, GRB 070129, GRB 160623A and GRB 221009A. 
We used four 3D extinction maps that exploit photometric data from different surveys and apply diverse algorithms for the 3D mapping of extinction, and compared the X-ray halo-derived distances with the local maxima in the 3D extinction density distribution. We found that in all GRBs we can find at least one local maximum in the 3D dust extinction map that is in agreement with the dust distance measured from X-ray rings. For GRBs with multiple X-ray rings, the dust distance measurements coincide  with at least 3 maxima in the extinction map for GRB 160623A, and 5 maxima for GRB 221009A. The agreement of these independent distance measurements shows that the methods used to create dust extinction maps may potentially be optimized by the X-ray halo observations from GRBs.
\end{abstract}

\begin{keywords}
X-rays: ISM – dust, extinction – gamma-ray burst: general
\end{keywords}

\section{Introduction}
The possibility of using X-ray scattering on interstellar dust grains to study the properties of dust such as its spatial distribution and the dust population, was pointed out early  by several authors \citep[e.g.,][]{overbeck1965, martin70}. The observations were limited by the imaging capabilities of the early 
X-ray telescopes, and the first dust halos were observed only in the eighties by the Einstein Observatory around bright Galactic sources  \citep{rolf1983, catura1983}. 
The theory of X-ray scattering from astrophysical sources was detailed in a number of works \citep[see e.g.,][]{mauchegorenstein86,mathis91,smith98,draine03,xiang11}. 
The observed intensity of the X-ray halo at different energies depends on the energy spectrum of the source,  column density of dust and the differential scattering cross section. Investigating the  energy- and time-dependence of scattering halos is thus crucial to infer the properties of grain sizes, chemical abundances, distances and spatial distribution of the dust layers \citep[e.g.][]{trumper73,mathis91,mir99,predehl00,draine03,costantini05}.
The differential cross section can be calculated using the exact Mie solution for scattering on spherical particles or adopting the Rayleigh-Gans approximation, which is valid above $\sim$ 2 keV 
\citep{mauchegorenstein86, mathis91, predehl95,smith98}. The X-ray scattering by nonspherical grains was calculated by e.g. \citet{draine06} who showed that substantial anisotropy of the X-ray halo may be expected for aligned interstellar grains and realistic size distributions. 

The search
for X-ray halos around bright Galactic X-ray sources was performed using different surveys, e.g. {\it ROSAT} by \cite{predehl95} or {\it Chandra} and {\it XMM-Newton} by \cite{valencicsmith2015}. Rings, or halos, were also detected around a plethora of magnetars \citep[e.g.][]{tiengo10, svirski11, mereghetti20}.

Gamma-ray bursts (GRBs), as impulsive bright X-ray events, offer a tool to infer the distance of the intervening dust when located behind sufficiently large Galactic column densities along the line of sight.
For short X-ray impulses scattered by individual dust clouds, the X-rays scattered at larger angles with respect to the line of sight will arrive at the observer with a time delay, and expanding rings will be formed.
To date, measurements of dust-layer distances and modelling of the energy-dependent radial profiles of X-ray halos have been performed only on a limited sample of observed halos surrounding GRB sources \citep{vaughan04,vaughan06,tiengo06,vianello07,pintore17,tiengo2023,vasilopoulos23, williams23}. The increasing importance of such observations was recently pointed out by \citet{nederlander20} who proposed X-ray halo observations as a tool for locating the electromagnetic counterparts to gravitational wave sources. 

The cross-section for scattering by dust increases rapidly with grain size and X-ray sources can contribute to the current constraints obtained from optical, ultraviolet and infrared observations, providing complementary information on the properties of large particle grains \citep[larger than a few $\mu$m,][]{mathis91}. The sky regions with detected halos around GRBs were studied in different energy bands: e.g. \citet{vaughan06} used X-ray 0.4 - 1.2 keV {\it ROSAT} all-sky survey data and the {\it IRAS} all-sky survey 100 $\mu$m map around GRB 050724, showing that the infrared (IR) dust emission and soft X-ray absorption were correlated and therefore caused by the same medium. On the contrary, the 21-cm map of atomic hydrogen (H {\small{I}}) of the region showed no correlation with these images, suggesting the comparatively lower density of H {\small{I}}. \citet{pintore17} 
measured distances of dust layers from X-ray observations of halo from GRB 160623A and compared them to a 3D map of interstellar reddening   \citep{green2015}.
They found high levels of extinction at several distances, with the largest extinction coinciding with the main dust layer identified in the X-ray data. The H {\small{I}} profile showed a peak possibly associated with the closest clouds identified in X-ray data, $\sim$0.5-1 kpc, and an extended region. The H$_2$ profile 
showed a peak at a different distance, $\sim$ 2 kpc \citep{pintore17}.

The brightest GRB of all times, GRB 221009A \citep{burns23}, occurred at low Galactic latitude ($b = 4\fdg3$) and produced more than 20 bright X-ray rings, observed by the
{\it Swift}
\citep{vasilopoulos23, williams23}, IXPE \citep{negro2023} and {\it XMM-Newton} \citep{tiengo2023}. In particular, 
\citet{tiengo2023} reported {\it XMM-Newton} observations of 20 rings around GRB 221009A, resulting from scattering on dust layers at distances from 300 pc to 18.6 kpc. They used the column density based on 3D extinction maps to estimate the GRB fluence, which allowed to constrain the prompt X-ray emission of the burst in the 0.5-5 keV energy band.

Similar studies have been performed for the dust scattering halos in the supernova remnant HESS J1731-347 \citep{landstorfer22}, where the dust distribution estimated from the 3D extinction maps from \cite{lallement19} was used to constrain the source distance, in addition to {\it Chandra} observations. These examples demonstrate that only combining different approaches helps to properly determine the distance of the scattering dust layers and to understand the physical process behind the observed X-ray halo intensity distribution. 

In this work for the first time the distances to dust clouds obtained using GRB X-ray observations are systematically compared with 3D maps of Galactic interstellar dust reconstructed through the 
tomographic inversion of extinction measurements toward stars with known distances. 
A large sample of reliably measured stellar data is required for this method to be successful. This became possible with the availability of massive stellar surveys, such as 2MASS \citep{Skrutskie2006}, ALLWISE \citep{Wright2010,Mainzer2011}, Pan-STARRS \citep{Chambers2016}, and the recent arrival of the \emph{Gaia} mission \citep{Gaia2016,Gaia2018,Gaia2021}.  Among a number of available 3D maps \citep[e.g.][]{sale18, chen19, rezaei20, hottier20, guo21}, we used a representative sample of them, done by \citet{green19}, \citet{leike20} and \citet{lallement19, lallement22}. They differ in the choice of the data used for extinction, stellar distance measurements, and applied inversion techniques.

The paper is organized as follows. In Sect.~\ref{sec:halos} we present the current status of the X-ray halo observations for a sample of gamma-ray bursts and the methods used to determine the distance to the scattering dust layers.   
Section~\ref{sec:3dmaps} describes the available 3D extinction maps of Galactic interstellar dust and the methods used to generate them. We present the case study of GRB 160623A, for which we show the extinction density distribution and compare it with the distances of dust layers along the line of sight obtained from X-ray data. The same method was applied to the whole sample of GRBs for which X-ray halos were observed. We discuss our results and possible discrepancies between the dust layer distances determined from these two methods in Sect.~\ref{sec:conc}.

\section{Determination of distances from X-ray halo observations}\label{sec:halos}
Dust scattering halos are nowadays often observed around bright X-ray objects, the number of which is increasing thanks to the imaging capabilities of the current X-ray instruments onboard {\it XMM-Newton}, {\it Chandra} and {\it Swift}. The basic process responsible for dust halos is the scattering of X-ray photons by
grains of the interstellar dust layers between us and the X-ray source. The scattering angles involved in this process are small, see e.g. \cite{draine03}. Scattered X-ray photons arrive with a certain delay, related to the
travelled path, with respect to the unscattered photons. Therefore, slow flux variations of the illuminating central X-ray object can be observed in changes of the dust halo flux. When the X-ray source is impulsive, as in the case of bursts, flares, or GRBs, the dust scattering halo is observed as an expanding ring. In the thin layer approximation for the intervening dust cloud, the angular radius $\theta(t)$ of the ring can be expressed as:

\begin{equation}
\theta(t) = \bigg[ \cfrac{2c}{d} \cfrac{(1-x)}{x} (t-\text{T}_0) \bigg]^{0.5},
\end{equation}

\noindent
where {\it x} = $d_\text{dust}/d$ ($d$ and $d_\text{dust}$ are the impulsive source and dust layer distances, respectively), {\it c} is the speed of light and T$_0$ is the time of the burst. This relation shows that, when $d$ is much larger than $d_\text{dust}$ (as it is the case for GRBs), it could be simplified as follows:

\begin{equation}
\theta(t) \approx \bigg[ \cfrac{2c (t-\text{T}_0)}{d_\text{dust}} \bigg]^{0.5},
\label{eq:ring}
\end{equation}
\noindent
removing the previous degeneracy between the source and the dust-layer distances. Such a case shows up for GRBs illuminating dust layers in our Galaxy. Once the ring and its expansion rate are measured, it is possible to univocally determine the dust layer distance. Therefore, this method can be used to map dust regions of our Galaxy with high precision.

\subsection{Gamma-ray bursts with observed X-ray halo}

Currently, expanding halos have been observed only for a handful of GRBs (see Table \ref{tab:present_measurements}). The individual distances to dust layers were determined using different methods for different GRBs. For the analysis of the X-ray halo around GRB 031203, \citet{vaughan04} used {\it XMM-Newton} EPIC MOS (0.7 - 2.5 keV) data and created a background-subtracted radial profile of counts from several time intervals of $\sim$ 6000 s duration. For this particular GRB, there were two peaks corresponding to the two expanding rings visible in the radial profiles; the change of radii as a function of time was found to be consistent with Eq.~\ref{eq:ring}. The halo spectrum can be extracted from the annular region. For GRB 050724 the spectral model used to fit the halo spectrum was an absorbed power law \citep{vaughan06}. As expected, the halo spectrum was found to be steeper than the GRB X-ray spectrum due to the strong dependence of the scattering cross-section on energy. 

\citet{tiengo06} proposed a new method to analyze time variations of the dust-scattering halos, based on the construction of the so-called dynamical image. It consists of a 3D histogram containing the counts number, their arrival position with respect to GRB, and the arrival time. In such representation, the expanding ring is visible as a linear regression whose slope is inversely proportional to the distance of the scattering layer \citep{tiengo06, vianello07}. For each detected count in such 3D histogram, the distance $d_i= 2c(t_i-T_0)/\theta_i^2$ is computed. The distribution of $d_i$ includes both the halo photons and the background counts. The dust scattering rings result in clear peaks superimposed on the background contribution, which, if homogeneous, is distributed as a power-law with index --2.
The peaks were fitted with Lorentzian curves centered at the scattering layer distance. In Table \ref{tab:grbs_with_halos}, we present the dust scattering distances determined for the sample of all GRBs for which the X-ray halo has been presently observed. We provide the coordinates, the redshift, the fluence, the duration of the event, and the derived distances to the dust scatterings layers, including the FWHM of fitted Lorentzian functions. 

\begin{table*}
\fontsize{8}{10}\selectfont
\caption{Gamma-ray bursts for which the time variable X-ray halo was observed and the dust layer distances were determined from  X-ray observations by {\it Swift} XRT/{\it XMM-Netwon}. References for distance measurements: \citet{tiengo06} (T06); \citet{vaughan04} (V04); \citet{vaughan06} (V06); \citet{vianello07} (V07); \citet{pintore17} (P17); \citet{tiengo2023} (T23). Fluences are reported for the lowest energy band available. For GRBs observed by Swift, fluences are measured  in the enegy band 15-25 keV, and duration T$_{90}$ is determined using {\it Swift} BAT (15-150 keV), see \citet{3batcatalog}. Values derived from the different energy bands/instruments are marked with ($^*$). For GRB 031203 we adopted  the values from {\it INTEGRAL} GRB catalog, where 20-200 keV energy  band  was used for fluence  and for T$_{90}$ \citep{bosnjak14, vianello09}. For GRB 160623A fluence is obtained by extrapolating the Konus-Wind spectrum in the 0.3-10 keV range. The duration T$_{90}$ for this burst was determined for {\it Fermi} GBM energy band. The duration of GRB 221009A is adopted from \citep{frederiks2023} and it was estimated in 80-320 keV. The fluence is calculated in the 15-150 keV for this burst \citep{krimm22}. FWHM refers to the width of the Lorentzian fitted in the distribution of distances derived from the dynamical image \citep{tiengo06}. For GRB 070129 and GRB 050724, the analysis was based on a different method and no FWHM was estimated. }
\label{tab:grbs_with_halos}
	\begin{tabular}{c c c  c c  c   c c c  c}
	    \hline
		GRB & $z$&  fluence  & T$_{90}$ & {\it l}& {\it b} & instrument & distance &FWHM & ref. \\
            & &  [ 10$^{-7}$ erg/cm$^2$]  & [s]  & [$^\circ$]  & [$^\circ$]   & &  [pc] & [pc] & \\
		\hline
		\hline
		031203 &0.105 & 10.6$^*$ &  19$^*$ & 256 & -5  & {\it XMM-Netwon}  & 870 $\pm$ 5 & 82 $\pm$ 16 & T06, V04 \\
		 && & &&  & & 1384 $\pm$ 9 & 240 $\pm$ 30 &  \\
        \hline
		050713A & &7.1 & 125   &112 & 19  & {\it XMM-Netwon}  & 364 $\pm$ 6  & 33 $\pm$ 15 & T06 \\
	    \hline
	    050724 &0.258 &2.1 & 99  &350.4 & 15.1   & {\it Swift} XRT  &  139$\pm$ 9 & - &V06 \\
	    \hline
		061019 && 5.1 &180  &181.7& 4.3  & {\it Swift} XRT  & 941 $\pm$ 45 & 427 $\pm$ 107 & V07 \\
		\hline
		070129 & 2.338 &6.6 &460 &157.2& -44.7 &  {\it Swift} XRT & 150 & -& V07 \\
		& & & &&&  &  290  &   &  \\
		\hline
		160623A & 0.367& 120$^*$ &  107.8$^*$ & 84.2  & -2.7 &   {\it XMM-Netwon}  & 528.1 $\pm$ 1.2  & 23.4 $\pm$ 3.3 & P17 \\
		&& && & &  & 679.2 $\pm$ 1.9 & 32.2 $\pm$ 5.7 & \\
		& && & & &  & 789.0 $\pm$ 2.8 & 75 $\pm$ 10 &  \\
		& &&  & &&  & 952 $\pm$ 5 & 116 $\pm$ 15 & \\
		& && & &  & & 1539 $\pm$ 20 &  106 $\pm$ 60 & \\
		& && & &  & & 5079 $\pm$ 64 & 1000 $\pm$ 400 & \\
		\hline
        221009A &  0.151  & 740$^*$ & 284$^*$ & 52.9 & 4.3  & {\it XMM-Newton} & 300 $\pm$ 2  & 62$\pm$ 10  & T23 \\
        & && & &  & & 406.3 $\pm$ 0.2 & 26.9 $\pm$ 0.7 & \\
        & && & &   & & 439.8 $\pm$ 0.5 & 14.6 $\pm$ 1.9  & \\
        & && & &  & & 475.2 $\pm$ 0.3 & 30.9 $\pm$ 0.9 &\\
        & && & &  & & 553.6 $\pm$ 0.3 & 27.7 $\pm$ 1.0 & \\
        & && & &   & & 695.4 $\pm$ 1.2 & 23.1 $\pm$ 3.7 & \\
        & && & &   & & 728.6 $\pm$ 1.1 & 42.7 $\pm$ 2.5 & \\
        & && & &   & & 1027.3 $\pm$ 5.2 & 38.1 $\pm$ 8.7 & \\
        & && & &   & & 1161.7 $\pm$ 2.5 & 99 $\pm$ 21 & \\
        & && & &   & & 1831 $\pm$ 13 & 121 $\pm$ 44 & \\
        & && & &   & & 1973 $\pm$ 10 & 141 $\pm$ 52 & \\
        & && & &   & & 2129 $\pm$ 5 & 135 $\pm$ 14 & \\
        & && & &   & & 2599 $\pm$ 5 & 164 $\pm$ 18 & \\
        & && & &   & & 3075.5 $\pm$ 7.4 & 309 $\pm$ 28 & \\
        \hline 
 
	\end{tabular}
    \label{tab:present_measurements}
\end{table*}

\section{3D maps of Galactic interstellar dust towards the GRBs}\label{sec:3dmaps}
The distances to dust clouds measured using X-ray data can be better understood by examining the distribution of the dust along the GRB direction to the observer. In this section we study the local increases in the dust distribution towards the GRB, as measured by 3D extinction maps, and compare their locations with the distances of dust clouds obtained from X-ray halos. We used four 3D extinction maps from \citet{green19}, \citet{lallement19,lallement22} and \citet{leike20}, hereafter G19, L19, L22, and Le20 map, respectively. They differ in the choice of input data, applied reconstruction techniques, and extent of the mapped volume in the Galaxy. In addition, the G19 map is based on the spherical coordinate system which voxelizes the sky into pencil beams centered at the Sun,  while L19, Le20 and L22 maps are based on the Cartesian coordinate system centered at the Sun. Their brief overview is given in the following subsection.

\subsection{3D dust extinction maps}
\label{Available_3D_maps}
The G19 map combines stellar photometry from \emph{Gaia} Data Release 2 (DR2), Pan-STARRS 1 and 2MASS for the extinction estimates towards the stars with \emph{Gaia} DR2 parallaxes for the stellar distances. The dust distribution is inferred along each sightline by taking into account a spatial prior that correlates nearby sightlines. Details of this technique are presented in \citet{green19}. The map gives the cumulative extinction along sightlines in 120 logarithmically spaced bins of distances from 63 pc to 63 kpc. It covers sightlines in the sky north of a declination of $-30^\circ$. The angular resolution varies between 3.4 to 13.7 arcmin depending on the sky region. The extinction is given in arbitrary units that can be converted to magnitude in different bands using the coefficients in Table 1 of \citet{green19}. We use the r-band magnitude ($A_r$, effectively a magnitude at 6170 \AA) of Pan-STARRS 1 survey. The map is publicly available at the website\footnote{\url{http://argonaut.skymaps.info}} and within the \texttt{Python} package \texttt{dustmaps} \citep{green18}.

The L19 map combines \emph{Gaia} DR2 and 2MASS photometric data with \emph{Gaia} DR2 parallaxes. The dust distribution is inferred by the tomographic inversion of extinction measurements using a regularized Bayesian hierarchical technique described in \citet{lallement19}. This technique takes into account the spatial correlation of structures and adapts the resulting map resolution to the availability of measurements within a given region. The map has a resolution of 25 pc for structures within 1 kpc from the Sun and up to 500 pc in a few regions more distant than 3 kpc from the Sun. It covers $6\times6\times0.8~{\rm kpc^3}$ volume around the Sun and is publicly available in VizieR\footnote{\url{http://cdsarc.u-strasbg.fr/viz-bin/qcat?J/A+A/625/A135}}. The map provides the extinction densities (or differential extinction, $\text{d}A/\text{d}r$) in ${\rm mag~pc^{-1}}$ with magnitude defined at wavelength of 5500 \AA$\, $ ($A_0$).

The L22 map is an updated version of the L19 map. It combines \emph{Gaia} Early Data Release 3 (EDR3) and 2MASS photometric data with \emph{Gaia} EDR3 parallaxes. The inversion technique and the computational volume used is the same as in the L19 map. The larger available sample of stars and better accuracy of the Gaia EDR3 data compared to DR2 improved contrast between the peak densities and void regions in the L22 map and increased distances at which the structures are reconstructed. The map provides extinction densities in ${\rm mag~pc^{-1}}$ with $A_0$, as in L19. In addition, this map provides error estimates based on measured photometric and parallax errors, on availability of the measurements within some region, and on the correlation length used in the computation. The map has a resolution of 25 pc and is publicly available on the EXPLORE website\footnote{\url{https://explore-platform.eu}}. 

The Le20 map combines \emph{Gaia} DR2, 2MASS, PANSTARRS, and ALLWISE photometric data and \emph{Gaia} DR2 parallaxes. The tomographic reconstruction is done on a smaller volume but at a higher resolution than G19, L19 and L22 maps using variational inference and Gaussian processes.  The map covers $0.74\times0.74\times0.54~{\rm kpc^3}$ volume around the Sun and has a resolution of 2 pc. The map provides extinction densities  in ${\rm mag~pc^{-1}}$ defined in the natural logarithmic units of the G-band magnitude ($A_G$), effectively at 6400 \AA. The map is publicly available at VizieR\footnote{\url{http://cdsarc.u-strasbg.fr/viz-bin/cat/J/A+A/639/A138}} and within \texttt{dustmaps}.

\subsection{Extracting 3D maps towards the GRB halos}
To compare positions of peak extinction densities corresponding to dust layers from these maps with measurements done by tracing the X-ray halos of GRBs, we extracted extinction density distributions along the line of sight of each GRB in the sample. In the next section, the methodology is described for the case of GRB 160623A, while for the other GRBs in the sample, results are presented in Sect.~\ref{ssec:others} and in Appendix \ref{AppA}.

\subsubsection{Case study: GRB 160623A }
We used GRB 160623A as a case study  for our methodology, as X-ray observations of this GRB using {\it XMM-Newton} in the 1-2 keV energy band clearly showed six distinct X-ray halos, corresponding to different locations of dust layers along the line of sight.

We used linear interpolation to extract the 3D extinction density from the L19 map in the direction of the GRB. For the L22 map, we used the G-Tomo app on the EXPLORE website, which queries the data for given coordinates and distances. Similarly, we used the \texttt{Python} package \texttt{Dustmaps} for Le20 and G19 maps. To get extinction density from cumulative extinction of the G19 map, we took a derivative on the output of the \texttt{Dustmaps} query. It was also multiplied with the corresponding coefficient ($2.617$) from Table 1 in G19 to get the values in the $r$-band of the Pan-STARRS 1 survey.

\begin{figure*}
    \centering
    \includegraphics[width=0.49\textwidth]{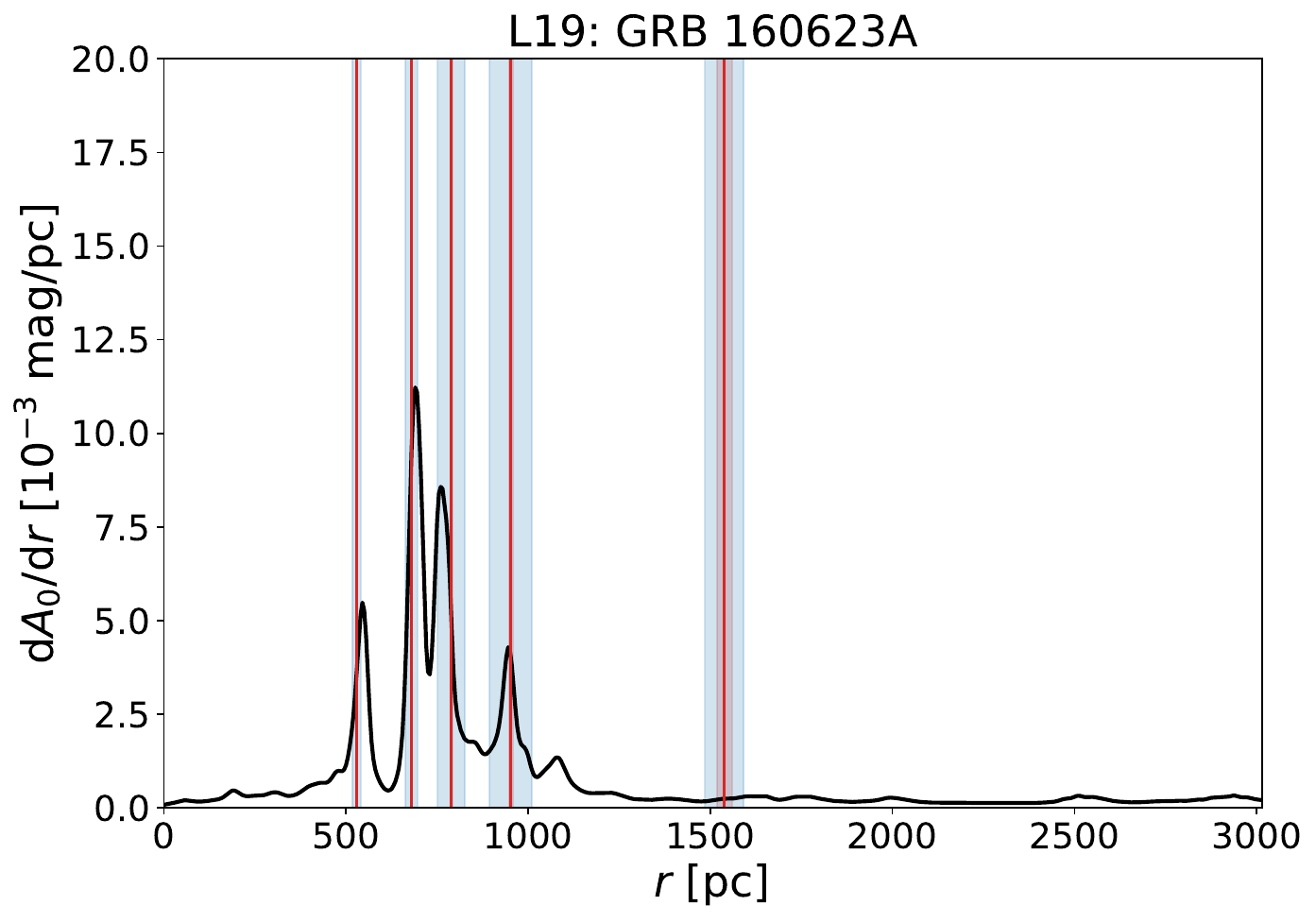}
    \includegraphics[width=0.49\textwidth]{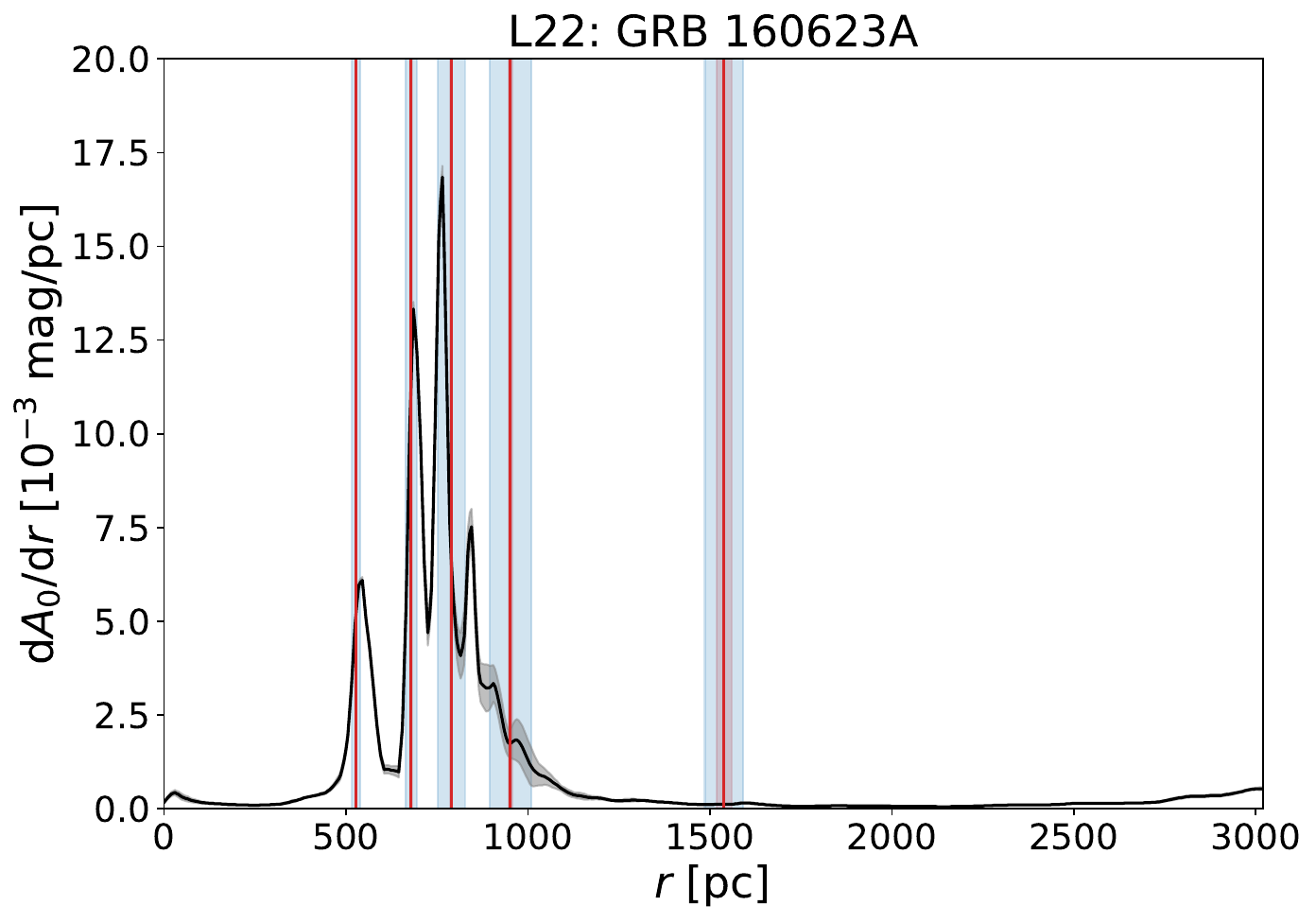}\\
    \includegraphics[width=0.49\textwidth]{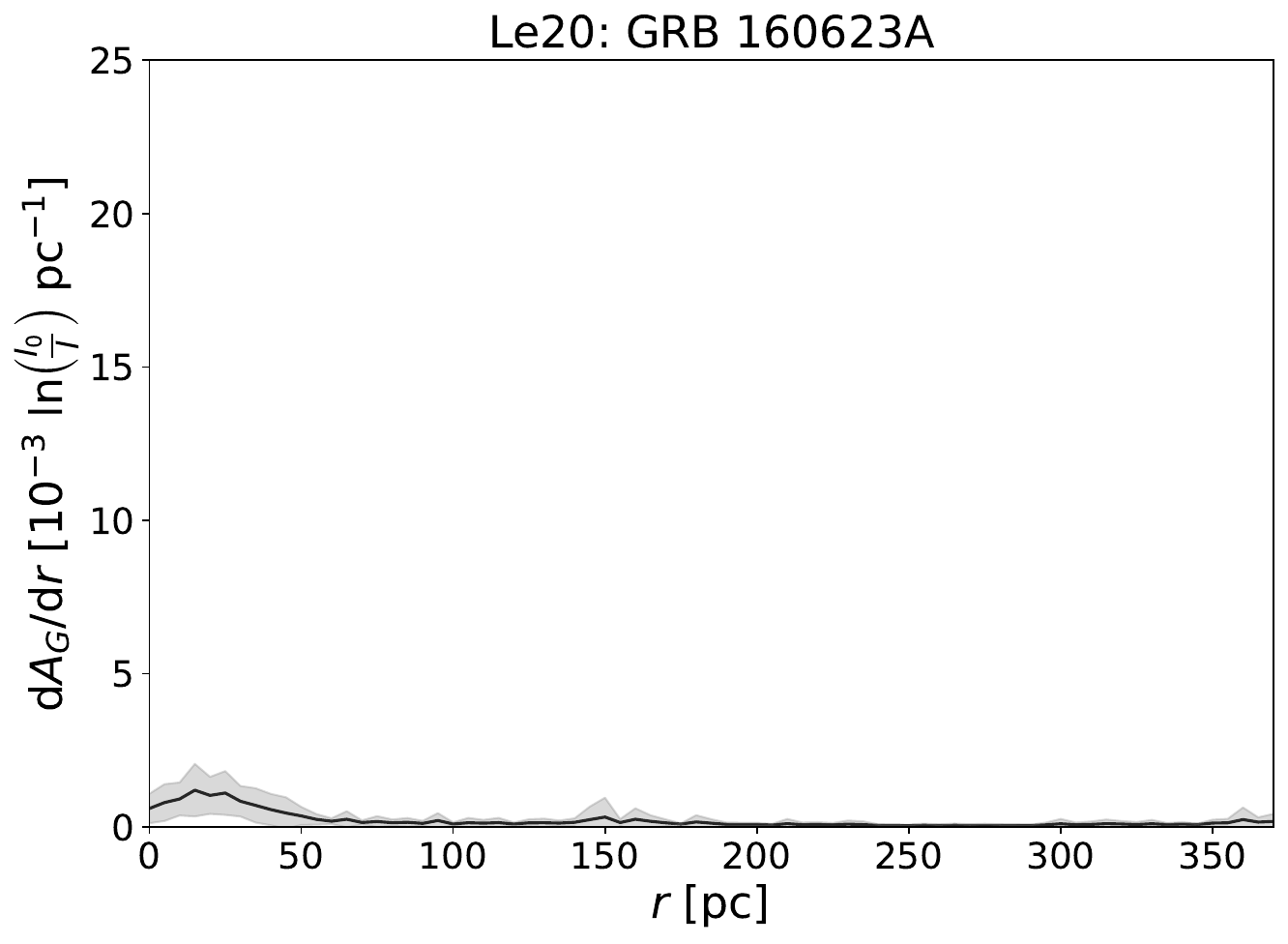}
    \includegraphics[width=0.49\textwidth]{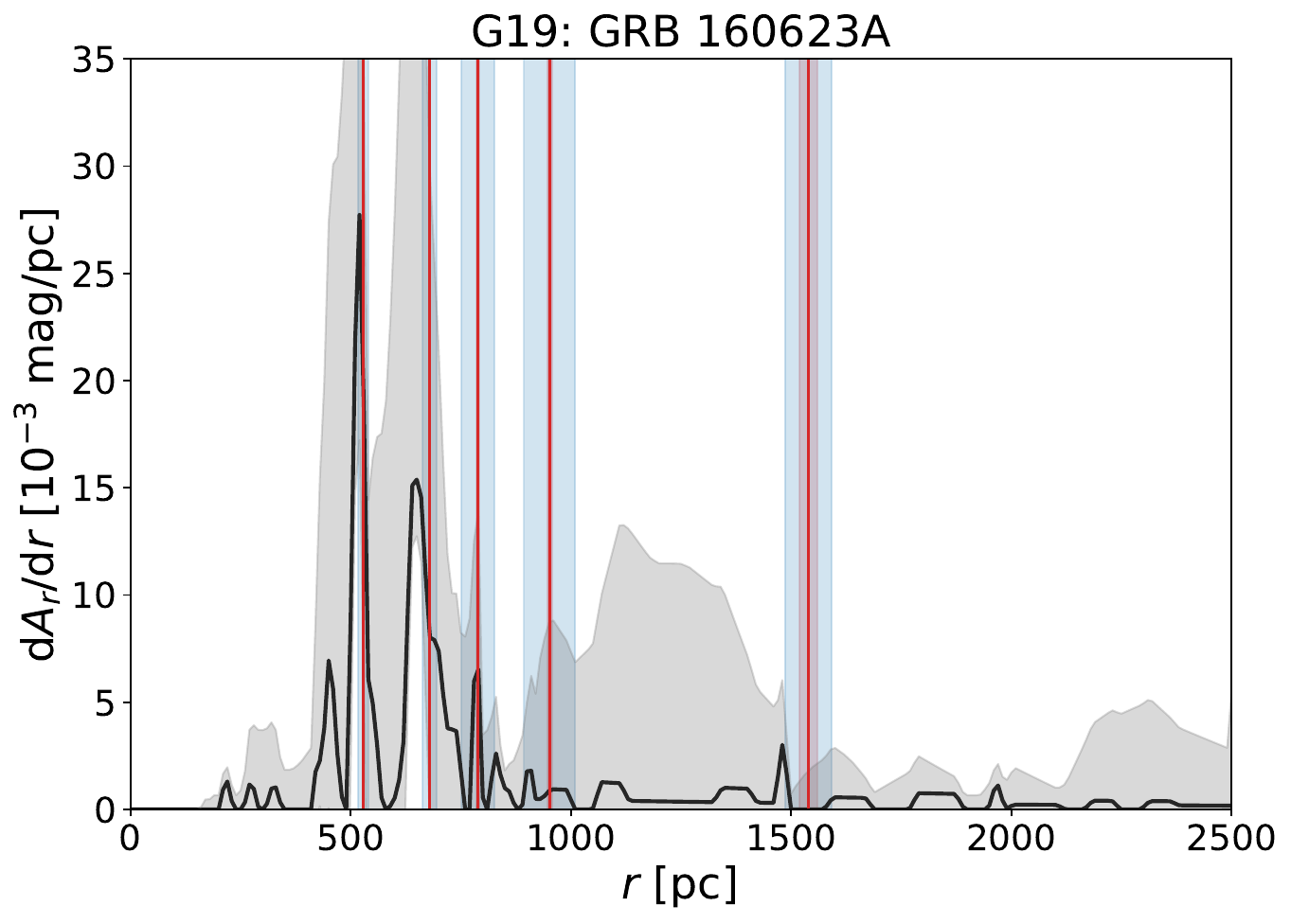}
    \caption{Extinction density distribution from L19 map (upper left), L22 (upper right), Le20 (lower left) and G19 (lower right) along the line of sight of GRB 160623A. Vertical red lines represent distances calculated from X-ray halos. Red shaded regions denote errors on these distances, while blue regions denote ranges covered by the FWHMs (see Table \ref{tab:grbs_with_halos}). When available (L22, Le20 and G19), the errors of extinction maps are shown as a grey-shaded region.}
  \label{fig:combo-160623A}
\end{figure*}

The results for each map are presented in Fig. \ref{fig:combo-160623A}. The errors on the distributions are available for L22, Le20 and G19 data. They are plotted as grey areas on the figures. The available coverage of distances for L19, L22 and Le20 maps limited us. The G19 dust distribution is presented for distances up to 2.5 kpc to focus on relevant distances for our analysis (see Table \ref{tab:grbs_with_halos}). Vertical red lines and red shaded areas mark distances and appropriate errors of X-ray halo measurements from Table \ref{tab:grbs_with_halos}. The blue shaded areas show the regions covered by the {\it FWHM} of the Lorentzian functions fitted in the $d_i$ distribution of counts from the dynamical image, see, e.g. \citet{tiengo06}.

\begin{figure*}
  \centering
    \includegraphics[width=0.48\textwidth]{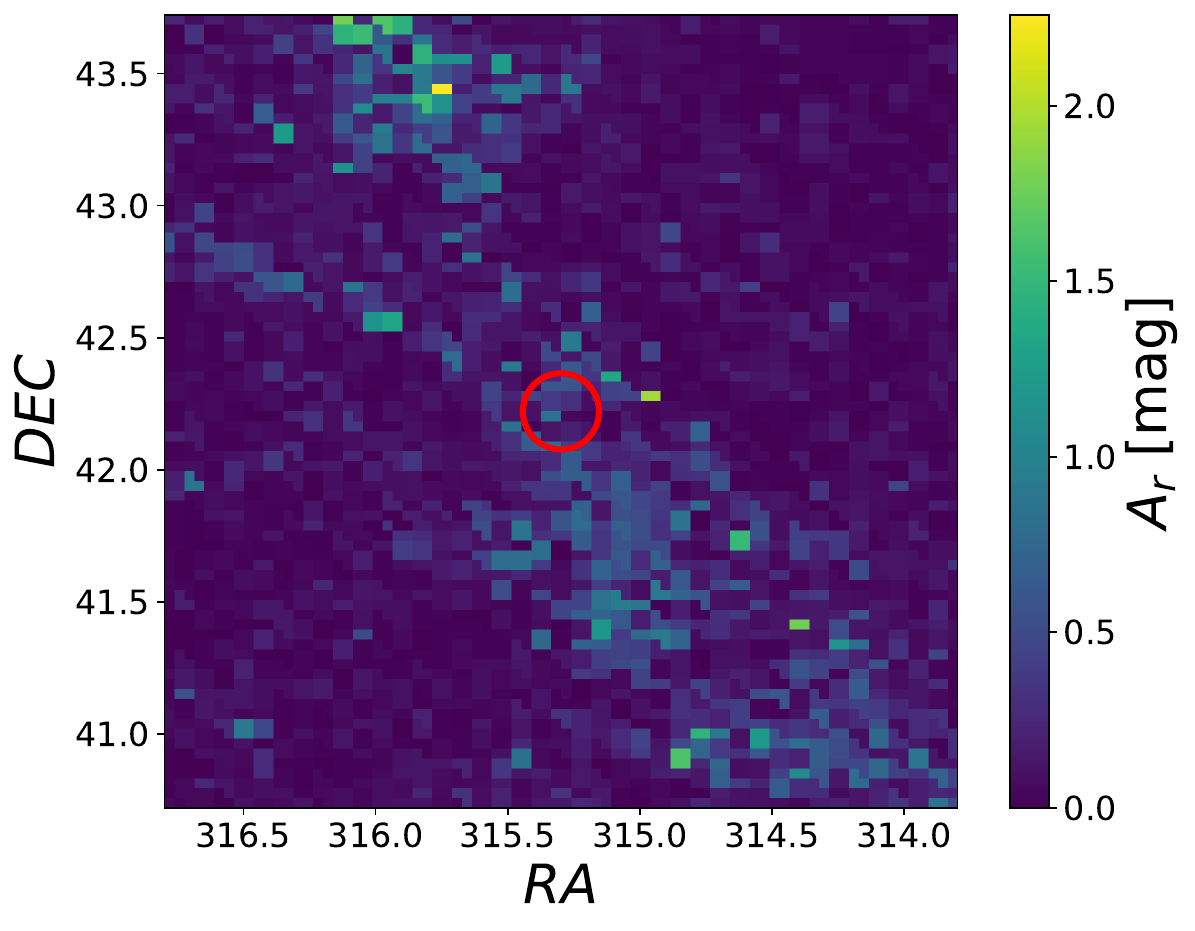}
    \includegraphics[width=0.48\textwidth]{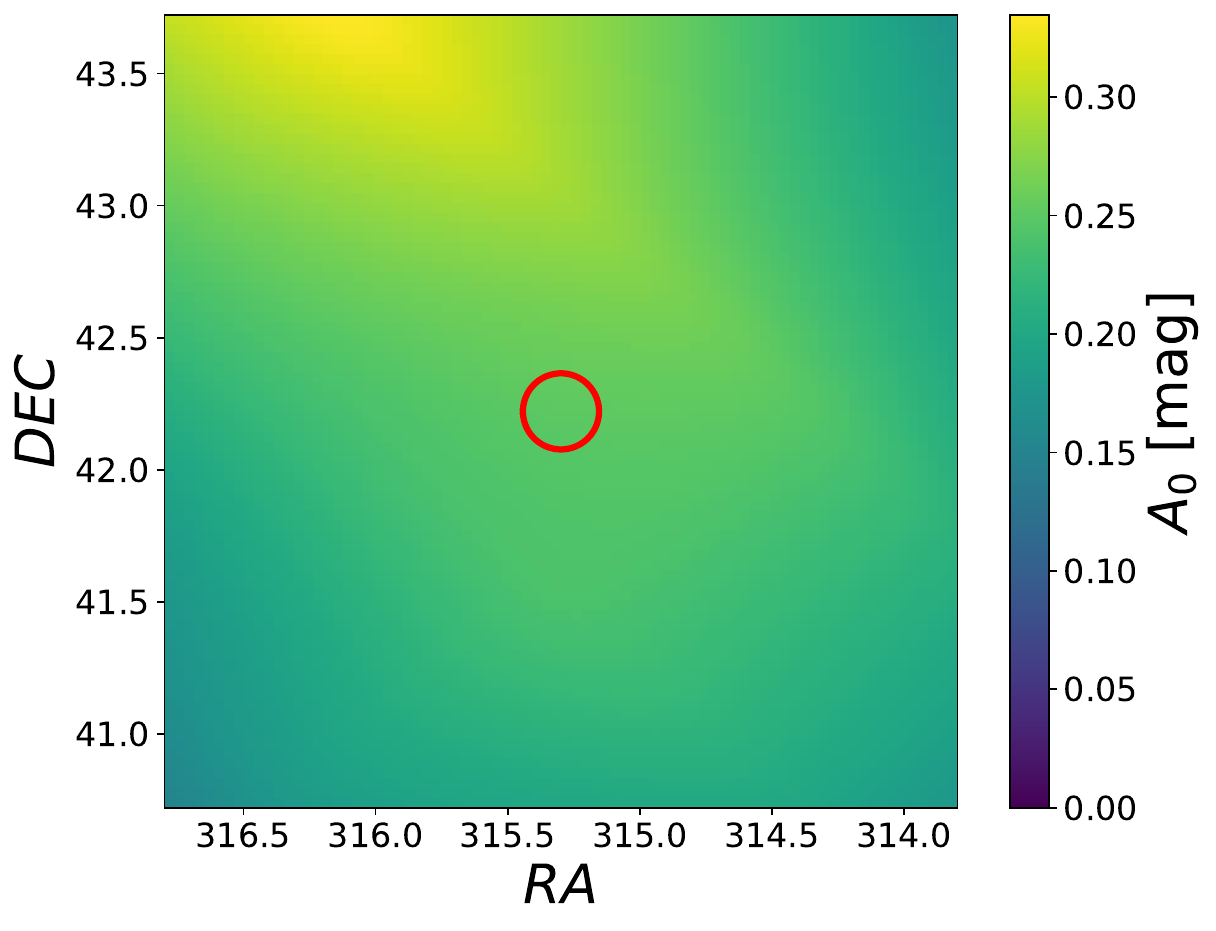}
  \caption{Integrated extinction from G19 (left) and L22 (right) around the distance measured from the X-ray halo for the nearest dust cloud in the case of GRB 160623A. The red circle represents the position of the observed halo.}
  \label{fig:combo-160623A-ring_1-2D-perpendicular}
\end{figure*}

The presented distributions of L19, L22 and Le20 maps are taken at the exact position of the GRB because the resolution of these maps is larger than the size of X-ray halos in the sky (of the order of a few arcmin). On the other hand, the G19 map has a resolution comparable to the halo sizes. We examined the surroundings of the GRB sky position at distances where the dust layer is measured from X-ray data. In Fig. \ref{fig:combo-160623A-ring_1-2D-perpendicular}, we show integrated extinction from L22 and G19 maps near GRB 160623A position around the distance ($\pm 20$ pc) of the closest dust cloud at 528.1 pc, as measured from X-ray data. The position of the ring is marked with a red circle in the plot. Due to high resolution, the G19 map shows small-scale fluctuations that are not present in the L19 map. Therefore, the simple extraction of one specific line of sight from this map is unreliable for this kind of study as close by sightlines (few arcmin) could significantly differ in extinction density distribution. Given this limitation, we are not using the G19 map for further analysis.

\begin{figure*}
  \centering
    \includegraphics[width=1\textwidth]{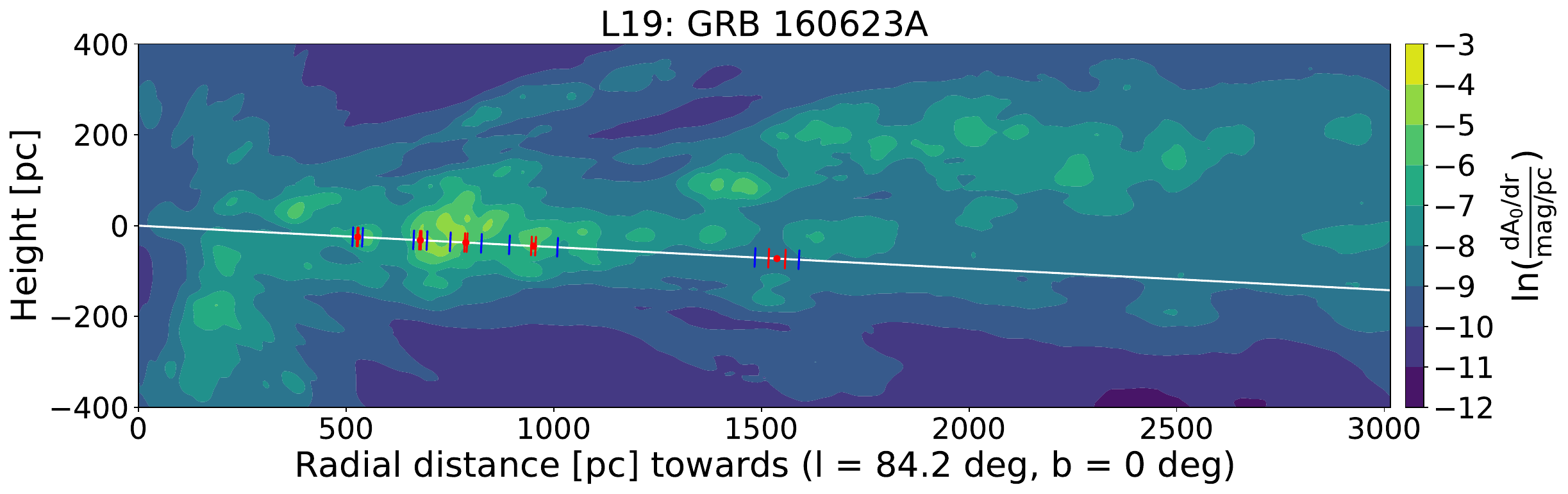}\\
    \includegraphics[width=1\textwidth]{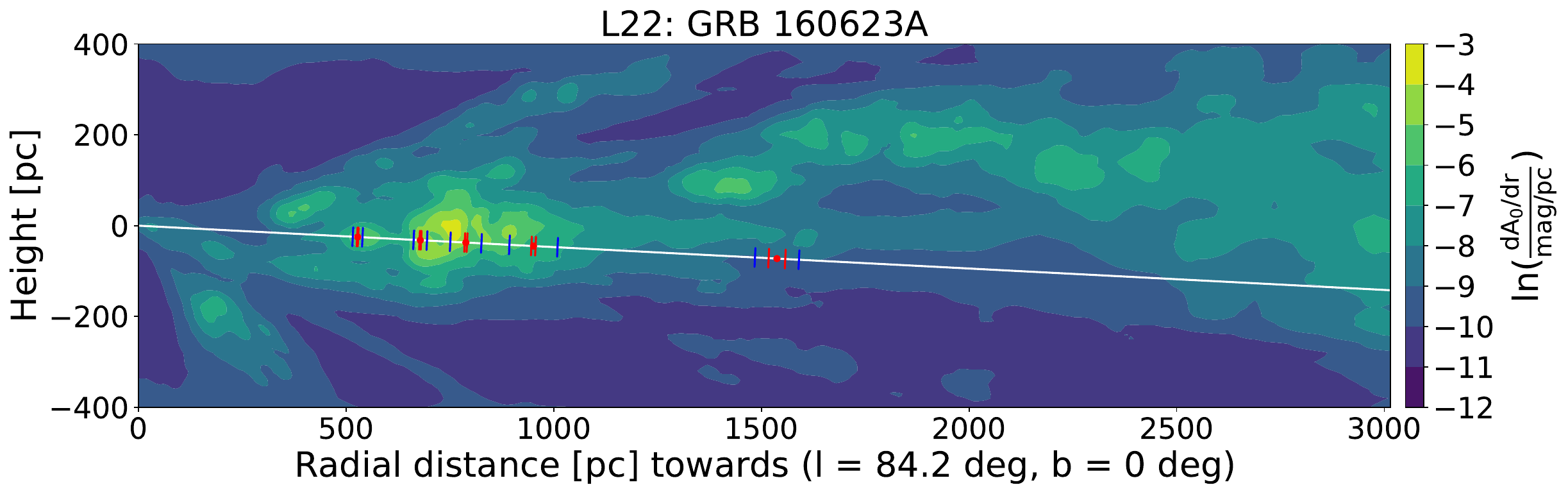}
  \caption{2D cut of extinction density cube from L19 map (up) and L22 (down) perpendicular to the Galactic plane  and in the direction of GRB 160623A. Height is measured with respect to the position of the Galactic plane. The white line represents the line of sight of the GRB, while red dots represent distances calculated from X-ray halos. Red perpendicular lines denote the errors on these distances, while the blue perpendicular lines denote ranges covered by the FWHMs  (see Table \ref{tab:grbs_with_halos}).}
  \label{fig:Lall2D-160623A}
\end{figure*}

To visualize the GRB surroundings along the line of sight, we made 2D cuts through L19 and L22 maps perpendicular to the Galactic plane (Fig. \ref{fig:Lall2D-160623A}). The line of sight toward the GRB is marked with a white line, while the corresponding measured distances of dust clouds are marked with red dots. The errors on distances and the {\it FWHM} of Lorentzians are given as red and blue lines perpendicular to the sightline, respectively.

The first three locations of the scattering dust regions are visible in L19 and L22 maps. The farthest two scattering layers are not identified as the resolution of these maps decreases at distances above $\sim$ 1 kpc. There is a good agreement between the positions of the maximum extinction in L19 distribution and the position of the first four scattering layers determined from the X-ray observations (see Table \ref{tab:grbs_with_halos}). The position of the fourth extinction maximum is instead shifted towards lower values in the L22 map. Moreover, the second extinction peak is higher than the third one in the L19 map, whereas the opposite is found in the L22 map. The reason for that can be understood from Fig. \ref{fig:Lall2D-160623A}. As mentioned in Sect.~\ref{Available_3D_maps}, the number of used sources for the L22 map is larger than for the L19 map, resulting in the increased contrast between the peak densities and void regions. Note that the Le20 data do not cover the distance to these dust layers and show only the structures within a distance of 100 pc in this case.

\subsubsection{Comparison of different distance measurements for the GRB sample}\label{ssec:others}

The same method was applied to the whole sample of GRBs with observed halos.
Results of our analysis are shown in
Appendix \ref{AppA} 
for all GRBs. We compared the scattering layers distances determined from the extinction density distribution for each GRB with those determined from the X-ray data.  In Table~\ref{tab:extinction_maxima} we give the positions of the local maxima in the extinction density distributions for L19, L22 and Le20 maps that are closest to the dust layers positions determined from X-ray studies. For the L22 map, the Gaussian was fitted to local maxima when there was no overlapping with nearby structures.

\begin{itemize}
    \item[--] {\it GRB 031203:} for this event an evolving halo around the source location was observed for the first time in a GRB. The {\it XMM-Newton} observations in 0.7 - 2.5 keV revealed two  expanding rings centred on the GRB. The rings were associated with X-ray scatterings on two distinct dust layers in the Galaxy, where the closest one was located at $\sim$ 880 pc, and the one further away at $\sim$ 1390 pc \citep{vaughan04}. As shown in Fig. \ref{fig:appendix_031203}, L19 and L22 maps show one distinct extinction density maximum corresponding to the smaller distance, while the Le20 map does not cover the large distances at which the dust layers were identified. The second layer distance determined from X-ray observations coincides with the elongated profile of a dust layer rather than a distinct peaked one. This is also visible in the 2D cut of the extinction density cube for GRB 031203 (last two rows in Fig. \ref{fig:appendix_031203}). Note that the FWHM of the Lorentzian fitted in the distance distribution from the dynamical image of GRB 031203 was rather large, 240 $\pm$ 30 pc.
    It corresponds to the wider line in the dynamical image, and this FWHM reflects the size of the region where the increase in extinction is seen in the maps 
    in Fig.~\ref{fig:appendix_031203}.\\
    
    \item[--] {\it GRB 050713A:} the scattering halo was not visible, but the dynamical image identified a dust-scattering layer. As there is only one clear maximum in the extinction density distribution visible in all maps (L19, L22 and Le20) shown in the first two rows in Fig. \ref{fig:appendix_050713A}, the agreement with the X-ray data is rather good. It is also visible in the 2D cut of the extinction density maps, shown in Fig. \ref{fig:appendix_050713A}, as a single maximum along the line of sight.\\
    
    \item[--] {\it GRB 050724:} As shown in Fig. \ref{fig:appendix_050724}, there is a structure in the extinction maps along the line of sight to this GRB, including the region of the Ophiuchus molecular cloud complex which is therefore a plausible site for scattering dust \citep{vaughan06}. The X-ray absorption and IR dust emission correlation pointed towards the same material in which these processes occurred. There is a good agreement with the highest maximum in the extinction density distribution from the L19 and L22 maps“, but the rest of the complex structure visible in the maps is not seen in the X-ray data. On the other hand,  Le20 has a higher resolution and better shows the distinction between the higher maximum at $\sim$150 pc and the much lower local maximum $\sim$280 pc.
    Considering the error, the dust distance measured from the X-ray data agrees with the first (and largest) extinction peak in the Le20 map. \\ 
    
    \item[--] {\it GRB 061019:} The dynamical image for this burst showed a rather wide line formed by the halo events \citep{vianello07}. This is reflected in the large FWHM value (107 pc) of the Lorentzian fitted in the distance distribution.
    Interestingly, the extinction density profile, shown in Fig. \ref{fig:appendix_061019}, shows
    several maxima distributed over an extensive range of distances, with the X-ray scattering layer distance closest to the position of the most significant maximum in L19 and L22 maps. The Le20 map shows only a shallow local maximum at $\sim$150 pc.\\
    
    \item[--] {\it GRB 070129:} The halo around this source had a relatively low number of counts, and therefore, the X-ray scattering layer distance was determined from the integral distribution of distances \citep{vianello07}. In this representation, the Lorentzian peaks become arctan profiles, and the distance to the scattering layer becomes the inflexion point. The results obtained using this method agree with the larger extinction maximum seen in L19, L22 and Le20 maps shown in Fig. \ref{fig:appendix_070129}. The distance to the closest dust layer determined from X-ray data is not visible in the L19 map, but there is an indication of an extended dusty region in the L22 and Le20 data. Note that the closer layer of dust was identified due to the statistical improvement of the fit when adding another inflexion point in the integral dust distance distribution.\\

    \item[--] {\it GRB 221009A:} The X-ray observations performed by {\it XMM-Newton} $\sim$2-5 days  after this exceptionally bright gamma-ray burst revealed 20 X-ray rings, produced by the dust layers at the distances ranging from 0.3 to 18.6 kpc \citep{tiengo2023}. The observations of dust-scattering rings were also reported using {\it Swift}/XRT \citep{williams23, vasilopoulos23} and IXPE \citep{negro2023} data. We used the data reported in \citet{tiengo2023} for our comparison, as the {\it XMM-Newton} observations could detect fainter X-ray rings and resolve multiple dust layers. We show the closest 14 peaks in the L19 and L22 maps and the first three peaks in the Le20 map in Fig. \ref{fig:appendix_221009A}. The maxima closest to 406.3, 475.2, 553.6, and 728.6 pc (which are measured using the X-ray observations) are easily identifiable in the L19 and L22 maps and coincide with the most prominent maxima in the extinction maps below 1 kpc. The closest maximum in the extinction maps corresponding to the layer at $\sim$240 pc is  poorly constrained in \cite{tiengo2023} (a Lorentzian centred at 300 pc, with a 62 pc width) because the corresponding ring was already mostly outside the instrument field of view during the first {\it XMM-Newton} observation.
    In 2D distributions from Fig. \ref{fig:appendix_221009A}, we see also that below 1 kpc there are several extended dust regions without distinct maxima. 
\end{itemize}

\begin{table}
\centering
\fontsize{8}{10}\selectfont
\caption{The local maxima in the extinction density distribution (along the line of sight of GRBs) closest to the distance determined from the X-ray studies. For the the L22 map, we fitted Gaussian functions to individual peaks that dodo not overlap with nearby structures and reported the obtained FWHMs (given in parentheses).}
\label{tab:extinction_maxima}
	\begin{tabular}{c c c c }
	    \hline
		GRB &  & maximum [pc]   &   \\
          & L19 map & L22 map & Le20 map \\
		\hline
		\hline
		031203 &  880 & 870 (52) & - \\
		       & 1375  & 1360 & -  \\
        \hline
		050713A & 355 & 365 & 360 \\
	    \hline
	    050724 & 160  & 130 (59) & 150 \\
	    \hline
		061019 & 1030 & 1005 (57)  & - \\
		\hline
		070129 & 295  & 305 (62) & 295\\
		\hline
		160623A & 545  & 545  & - \\
		& 690 & 685 & - \\
		& 760 & 765 (42) & -    \\
		& 945 & - & -   \\
	
		% & \textbf{970.0 (905.0)?} & \textbf{8.0 (47.0)?} &  \\
            \hline
            221009A & 230 & 250  & 230\\
            & 415 & 415 (43) & 400 \\
%            & - & - & -  \\
            & 465  & 465 (66) & -\\
            & 575 & 555 (47) & - \\
%            & - & - & -  \\ 
            & 725 & 735 (47) & - \\
        \hline
	
	\end{tabular}
\end{table}

\begin{figure}
  \centering
    \includegraphics[width=0.49\textwidth]{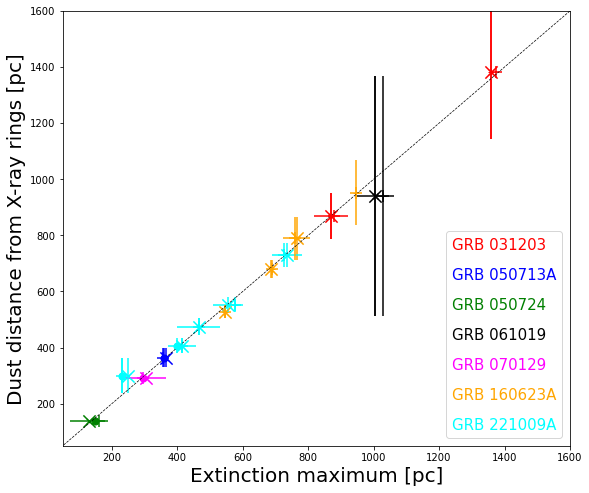}
  \caption{The distances measured from X-ray halos observations in comparison with the local maxima in the extinction maps. The crosses refer to L22 extinction map, the plus symbols refer to L19 map, and the circles to Le20 map. If the clouds are at distances $\gtrsim$ 400 pc, the Le20 data cannot be used (e.g. for GRB 031203, GRB 061019 and GRB 160623A).  For L22 map data, the errors of local maxima are estimated by fitting the Gaussian functions to individual peaks when the peaks are identifiable, see Table \ref{tab:extinction_maxima}. The dashed line shows the values for which these distances are equal (see Table \ref{tab:grbs_with_halos}). }
  \label{fig:results}
\end{figure}

\section{Conclusions}
\label{sec:conc}
The method proposed by \citet{tiengo06} to determine the distances to the dust scattering layers based on the dynamical image, in which each count is binned according to its arrival time and distance from burst (Eq. \ref{eq:ring}), allows to create the distribution of scattering layer distances.  The fit of the Lorentzian functions superimposed on the power law representing the background allows to determine the  distance of the scattering layers. 
The width of the Lorentzian peaks in the distance distribution is determined by the instrumental PSF (resulting in broader peaks for smaller rings), the GRB duration (which is relevant only at early times for sufficiently long GRBs) and the distribution of dust along the line of sight. This last effect can be either due to a single (geometrically) thick cloud or the combination of more clouds close to each other. In the latter case, since different distances imply a different expansion rate, two nearby clouds could appear as an unresolved peak at early times and then be resolved into two separate peaks in later observations (or in observations with better PSF or counting statistics). For example, in the {\it XMM-Newton} observation of GRB 221009A one can clearly distinguish two peaks at 698 and 729 pc \citep{tiengo2023} which appeared as a single peak in {\it Swift/XRT} observations which had poorer statistics \citep{vasilopoulos23}. Also, in the case of GRB 061019, \cite{vianello07} studied the width of the peak through simulations and found evidence for a significant intrinsic cloud width.

The extinction density distribution from three different extinction maps was extracted along the line of sight of each GRB for which the time-expanding halo is presently observed (Table \ref{tab:grbs_with_halos}). We show the comparison of distances derived using the X-ray halos with distances of dust regions from the individual extinction maps in Fig.~\ref{fig:results}. The number of dust layers that we can constrain is a function of fluence (Table \ref{tab:grbs_with_halos}), and dust layer density. Therefore, the  extinction maps and the X-ray observed distances are not always in accordance: the fainter is the GRB and less dense is the cloud, the more difficult is to constrain the position. In all GRBs that we examined, we found at least one local maximum in the 3D dust extinction maps that is in agreement with the dust distance measured from X-ray rings. When multiple rings were detected for a GRB, the dust distance measurements coincide with 4 (3) maxima in L19 (L22) map for the case of GRB 160623, and 5 maxima (in L19 and L22 maps) for GRB 221009A. We fitted a linear function to points corresponding to individual maxima in the extinction maps to check their agreement with the X-ray halo measurements. The fit to L22 data results in slope (1.02 $\pm$ 0.03), showing a good agreement of the two independent distance measurements. For the errors in dust distance, we used the FWHM of Lorentzian functions reported in Table \ref{tab:grbs_with_halos}, as it better captures the region in which the scatterings occur in case of the extended scattering regions. The errors for the extinction maxima were estimated for L22 map: we fitted Gaussian function to individual peaks when the peaks were identifiable (Table \ref{tab:extinction_maxima}). 

When individual dust layers are clearly separated, the distance measurements  from the X-ray data are in good agreement with the local maxima in the extinction density distribution.  This is clearly seen in the case of GRB 050713A. When there is no clear local maximum along the line of sight towards a GRB (see the 2D cuts of extinction density cube perpendicular to the Galactic plane along the line of sight towards the GRBs, Figs. \ref{fig:Lall2D-160623A}, \ref{fig:appendix_031203}-\ref{fig:appendix_221009A} ), but only extended regions where extinction occurs (e.g. in GRB 061019 or GRB 050724), we do not find clear correspondence with X-ray observations. If the distance to X-ray resolved dust rings is of the same order of magnitude as the resolution of the maps ($\sim$ 25 pc), it is not possible to capture two separate maxima in the dust extinction profile driven by the sparsity of the starlight data in a given direction.

Observations of X-ray halos can benefit from the study of dust extinction by providing information on the location and morphology of the scattering layers. Vice-versa, our comparison suggests that the method applied to create different dust extinction maps such as L19, L22 and Le20, could be potentially optimized by the use of X-ray halo observations from GRBs, as an independent distance measurement of dust layers in the Galaxy.

\section*{Acknowledgements}
We thank the anonymous referee for reviewing our manuscript and Rosine Lallement for her help with the G-TOMO module of the EXPLORE project. Ž.B. and V.J. acknowledge support by the Croatian Science Foundation for a project IP-2018-01-2889 (LowFreqCRO). A.B. acknowledges support from the European Research Council through the Advanced Grant MIST (FP7/2017-2022, No.742719). This research has used data, tools or materials developed as part of the EXPLORE project that has received funding from the European Union’s Horizon 2020 research and innovation programme under grant agreement No 101004214.

\section*{Data Availability}
All of the data underlying this article are already publicly
available from  {\url{http://argonaut.skymaps.info}, \url{http://cdsarc.u-strasbg.fr/viz-bin/qcat?J/A+A/625/A135}, \url{https://explore-platform.eu}, \url{http://cdsarc.u-strasbg.fr/viz-bin/cat/J/A+A/639/A138}}.
 
%%%%%%%%%%%%%%%%%%%% REFERENCES %%%%%%%%%%%%%%%%%%

\bibliographystyle{mnras}
\bibliography{reflist} % if your bibtex file is called example.bib
%%%%%%%%%%%%%%%%%%%%%%%%%%%%%%%%%%%%%%%%%%%%%%%%%%
%%%%%%%%%%%%%%%%% APPENDICES %%%%%%%%%%%%%%%%%%%%%
\appendix 
\onecolumn
\section{Extinction density distributions for GRB sample} \label{AppA}
\centering
\includegraphics[width=0.40\textwidth]{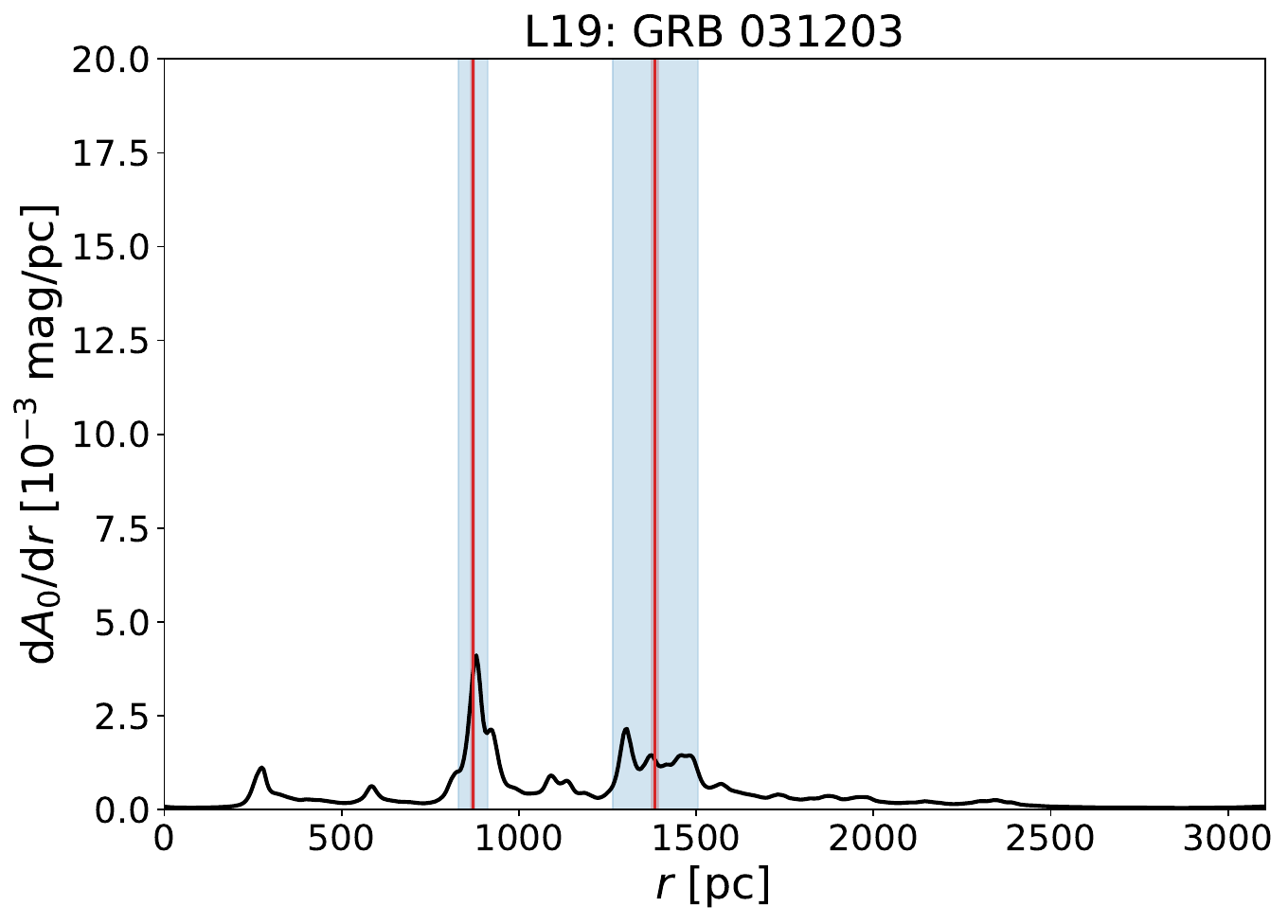}
\includegraphics[width=0.40\textwidth]{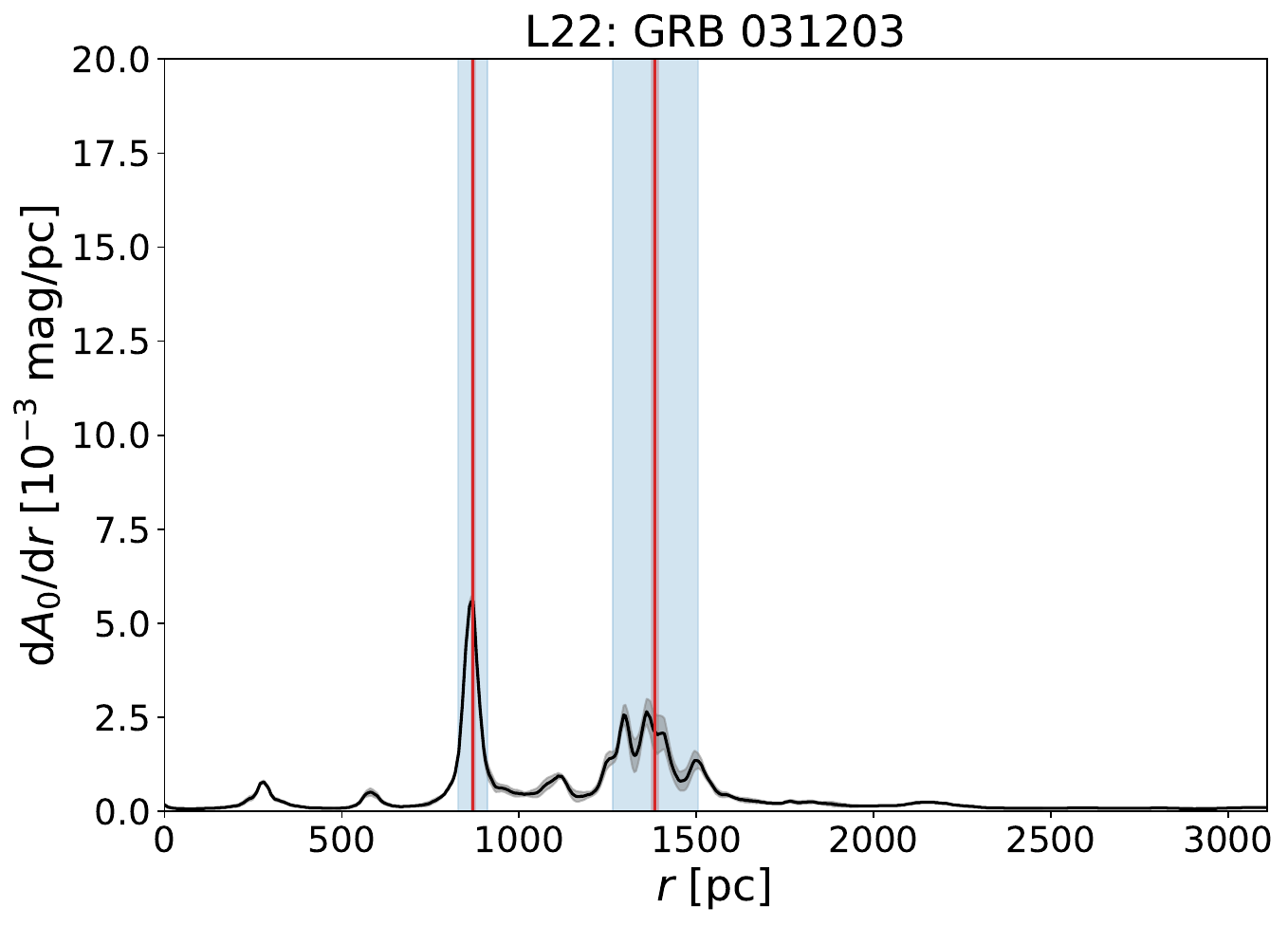}
\includegraphics[width=0.40\textwidth]{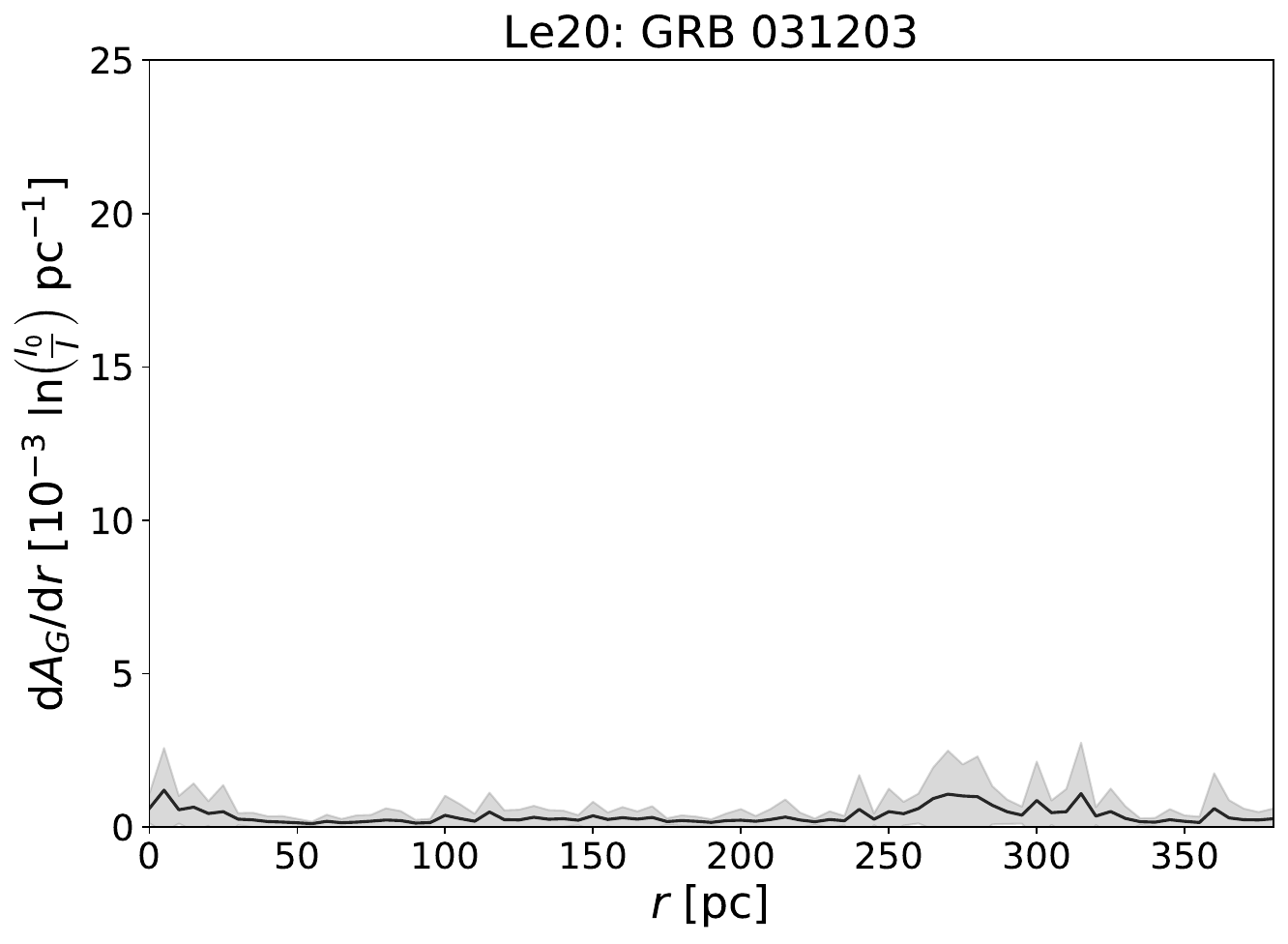}
\includegraphics[width=1\textwidth]{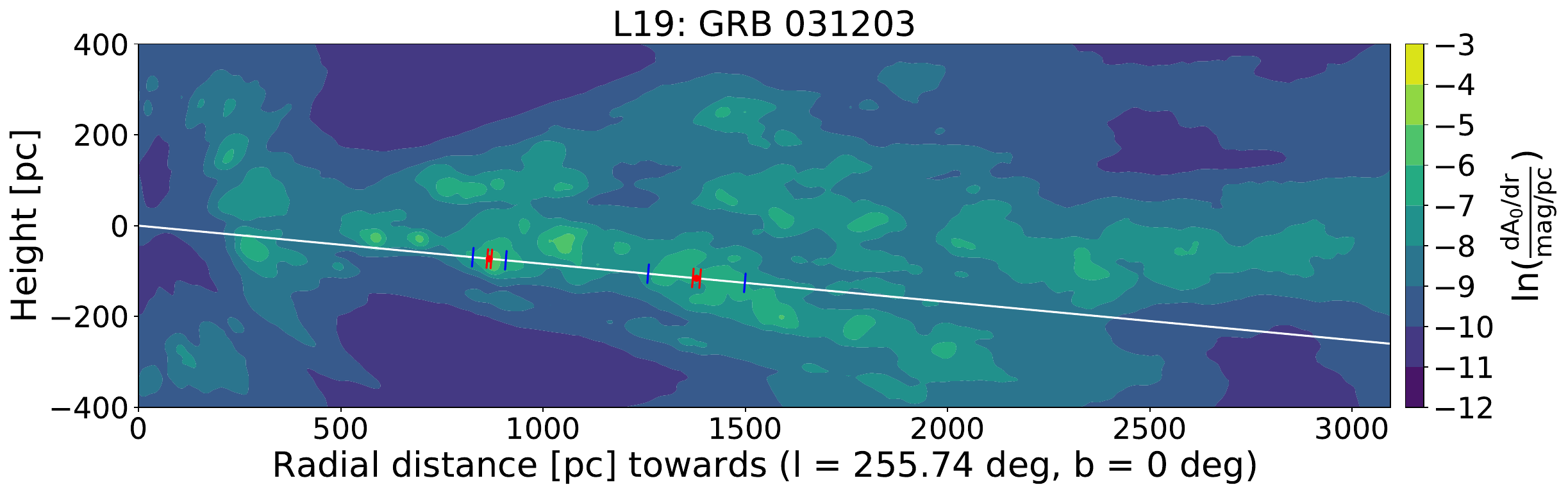}
\includegraphics[width=1\textwidth]{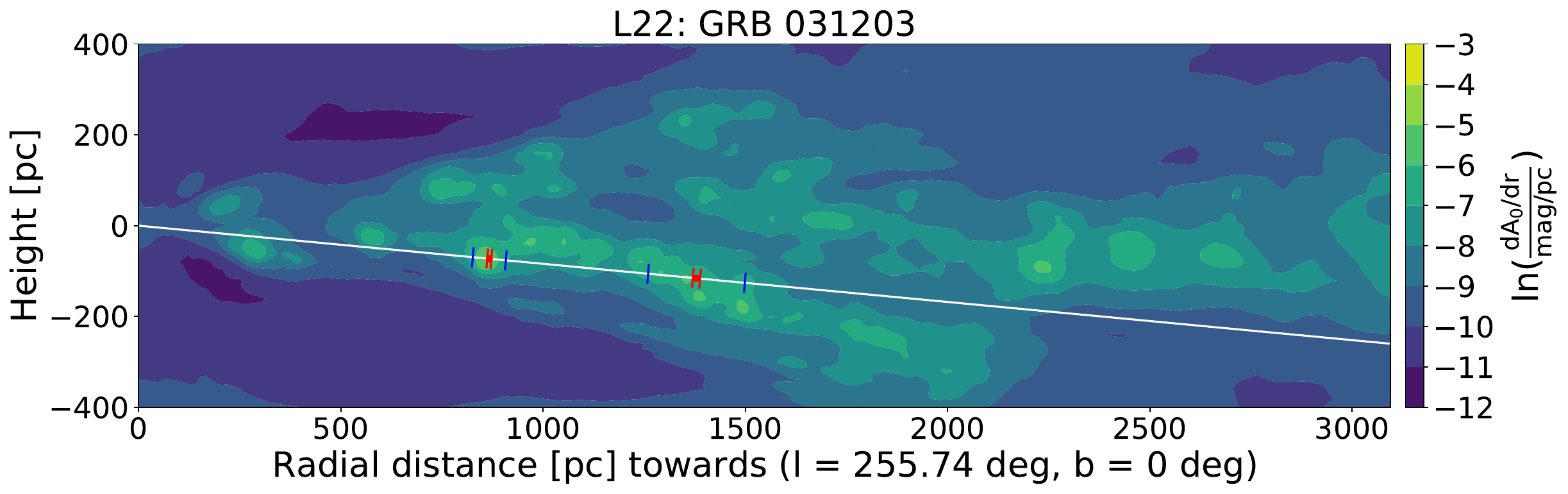}
\captionof{figure}{GRB 031203. First and second row same as in Fig.~\ref{fig:combo-160623A}, without G19 map. Last two rows same as in Fig.~\ref{fig:Lall2D-160623A}.}
\label{fig:appendix_031203}

\includegraphics[width=0.40\textwidth]{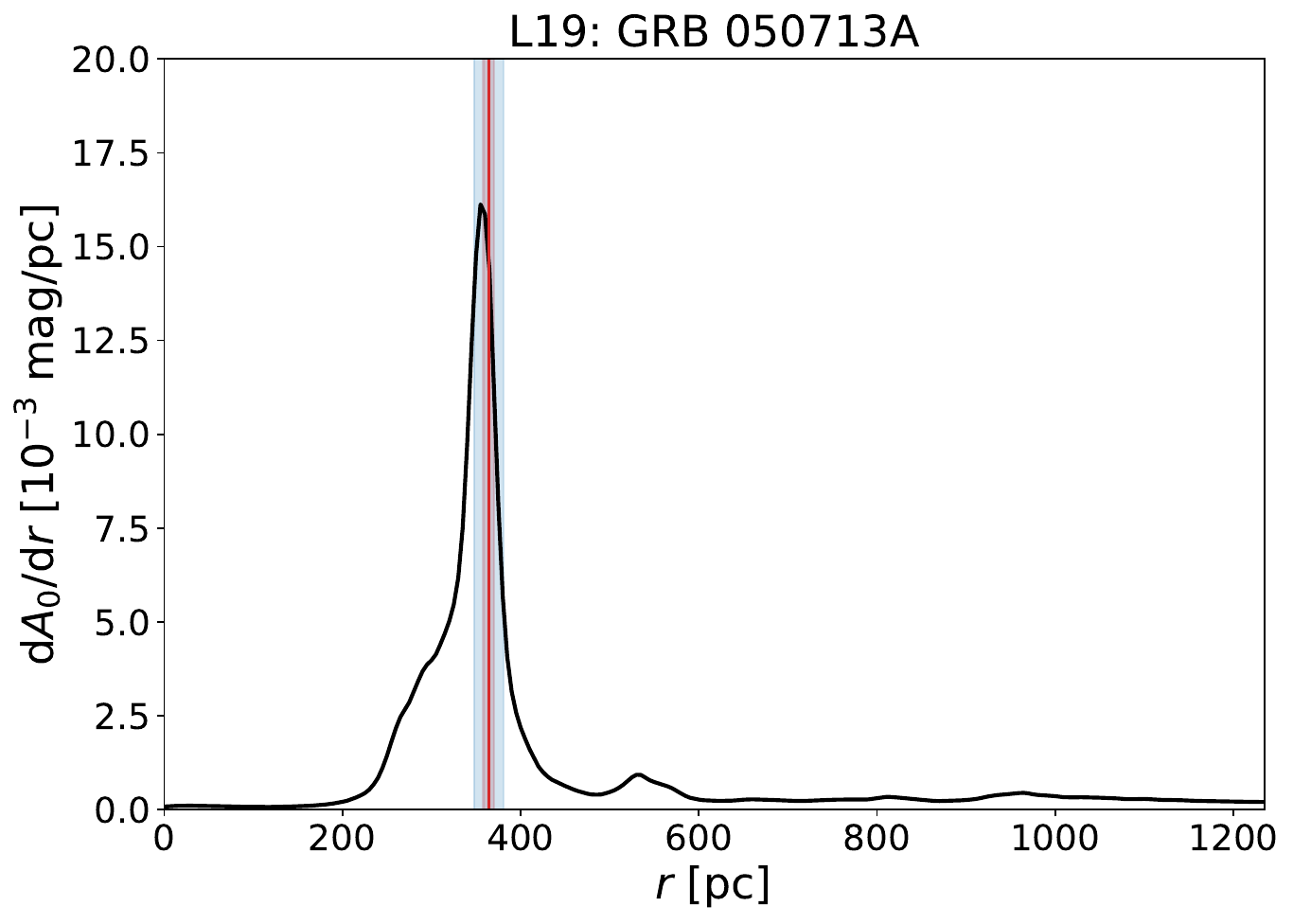}
\includegraphics[width=0.40\textwidth]{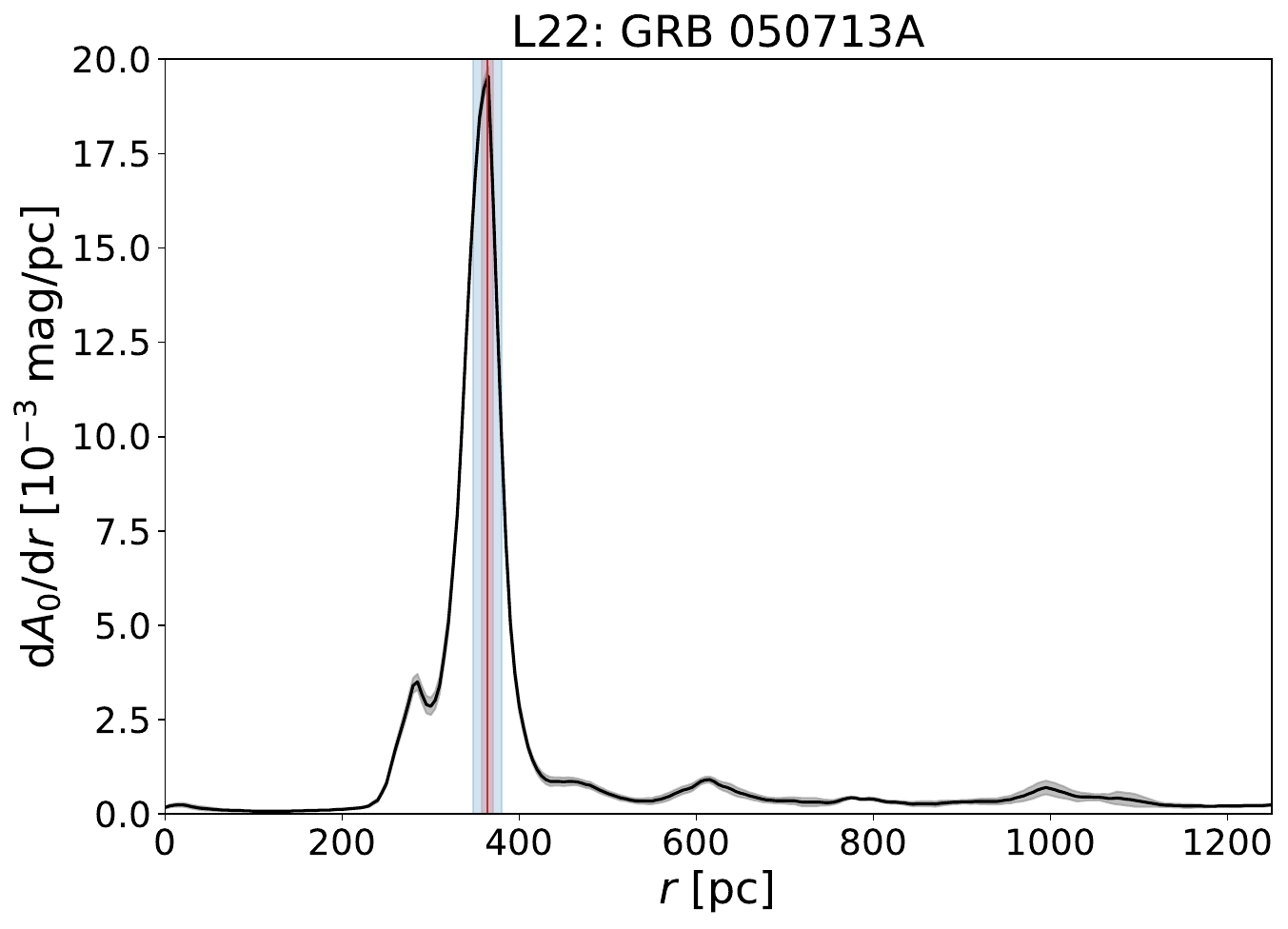}
\includegraphics[width=0.40\textwidth]{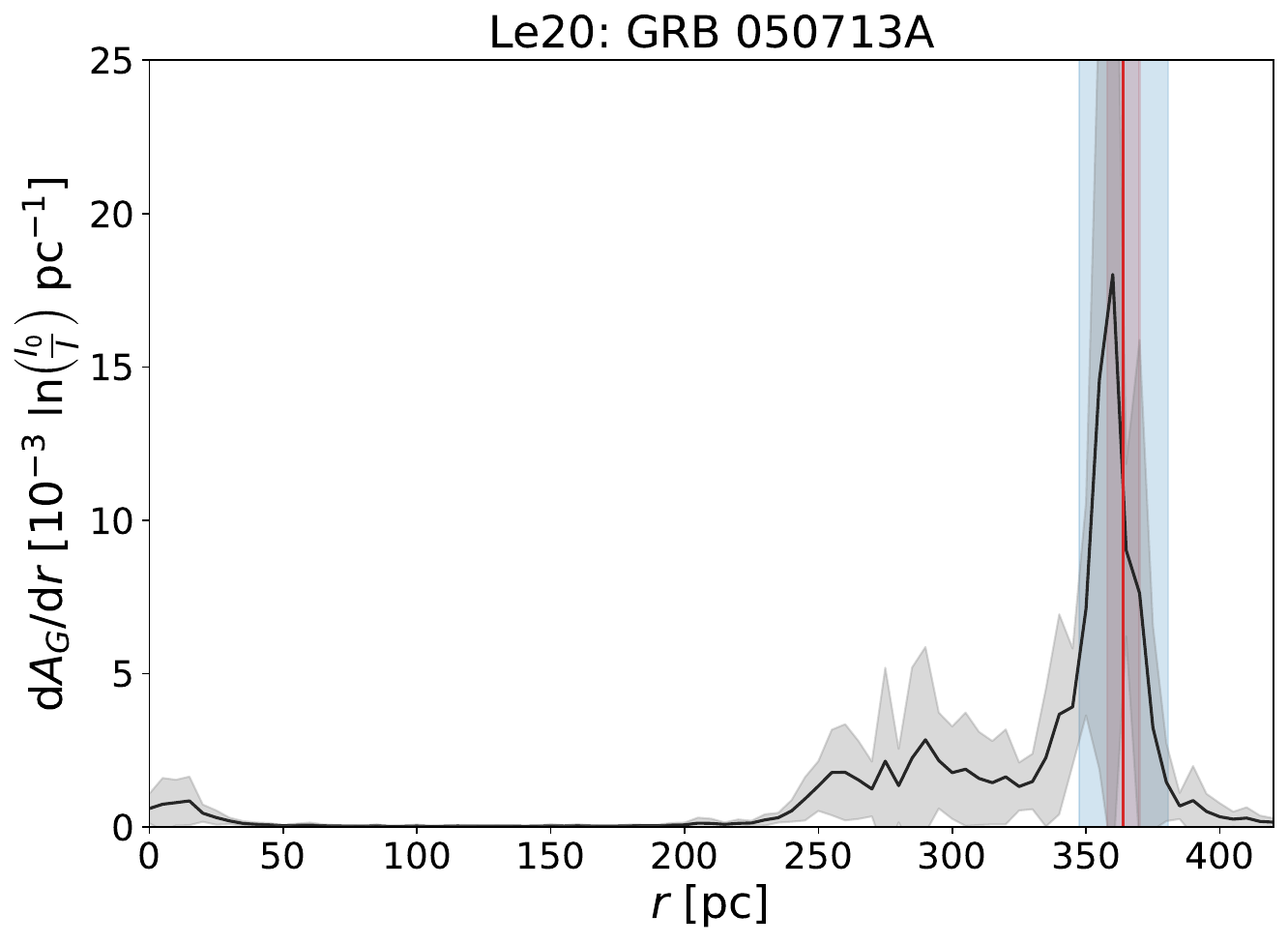}
\includegraphics[width=1\textwidth, height=4.8cm] {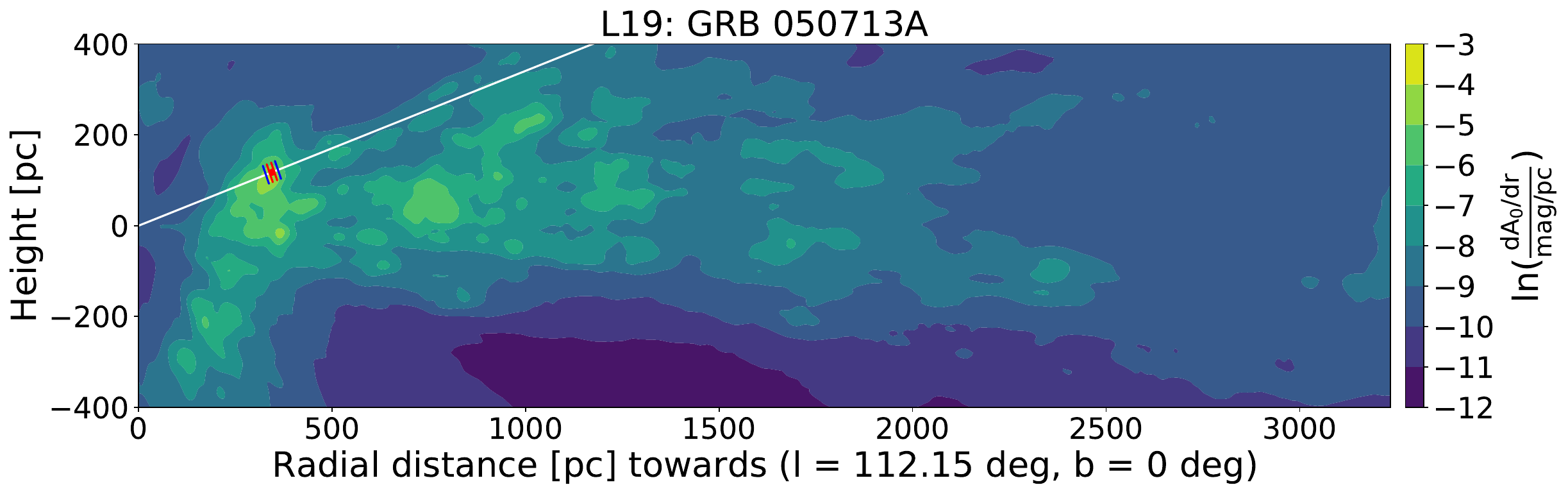}
\includegraphics[width=1\textwidth, height=4.8cm]{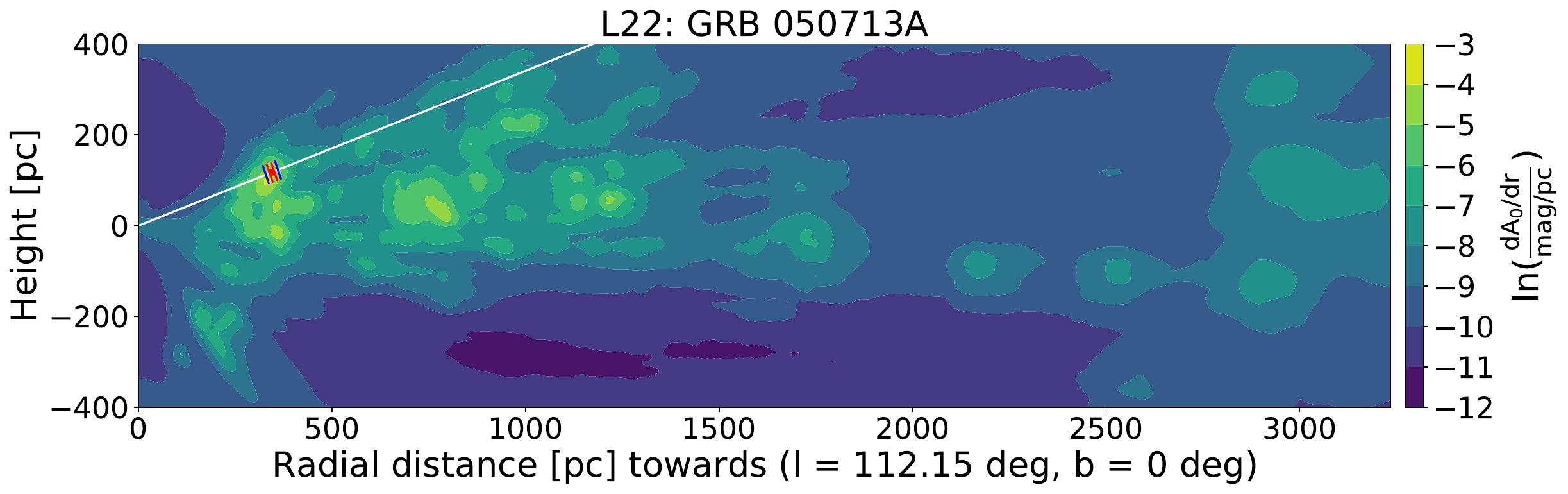}
\captionof{figure}{GRB 050713A. First and second row same as in Fig. \ref{fig:combo-160623A}, without G19 map. Last two rows same as in Fig. \ref{fig:Lall2D-160623A}.}
\label{fig:appendix_050713A}

\includegraphics[width=0.40\textwidth]{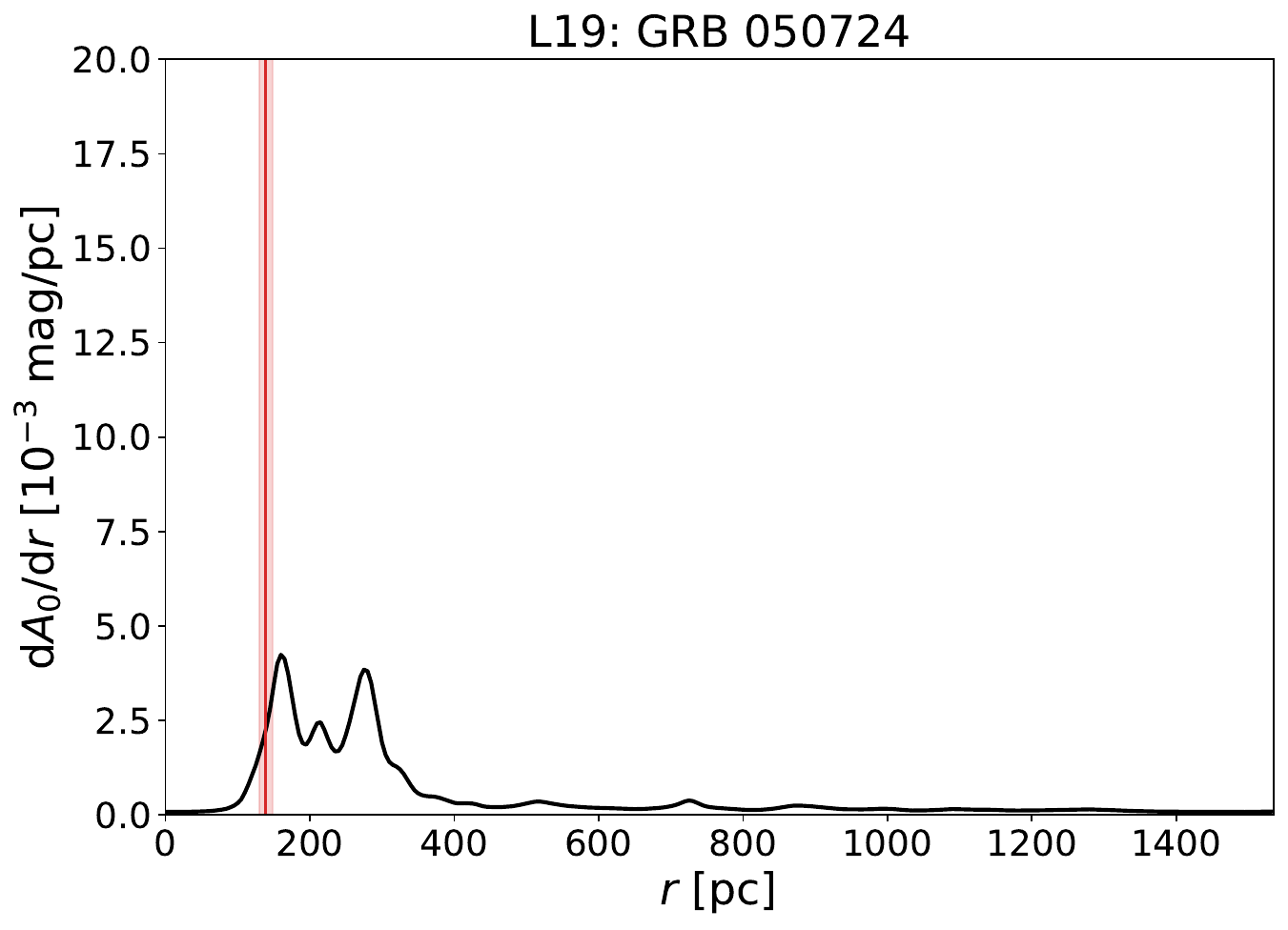}
\includegraphics[width=0.40\textwidth]{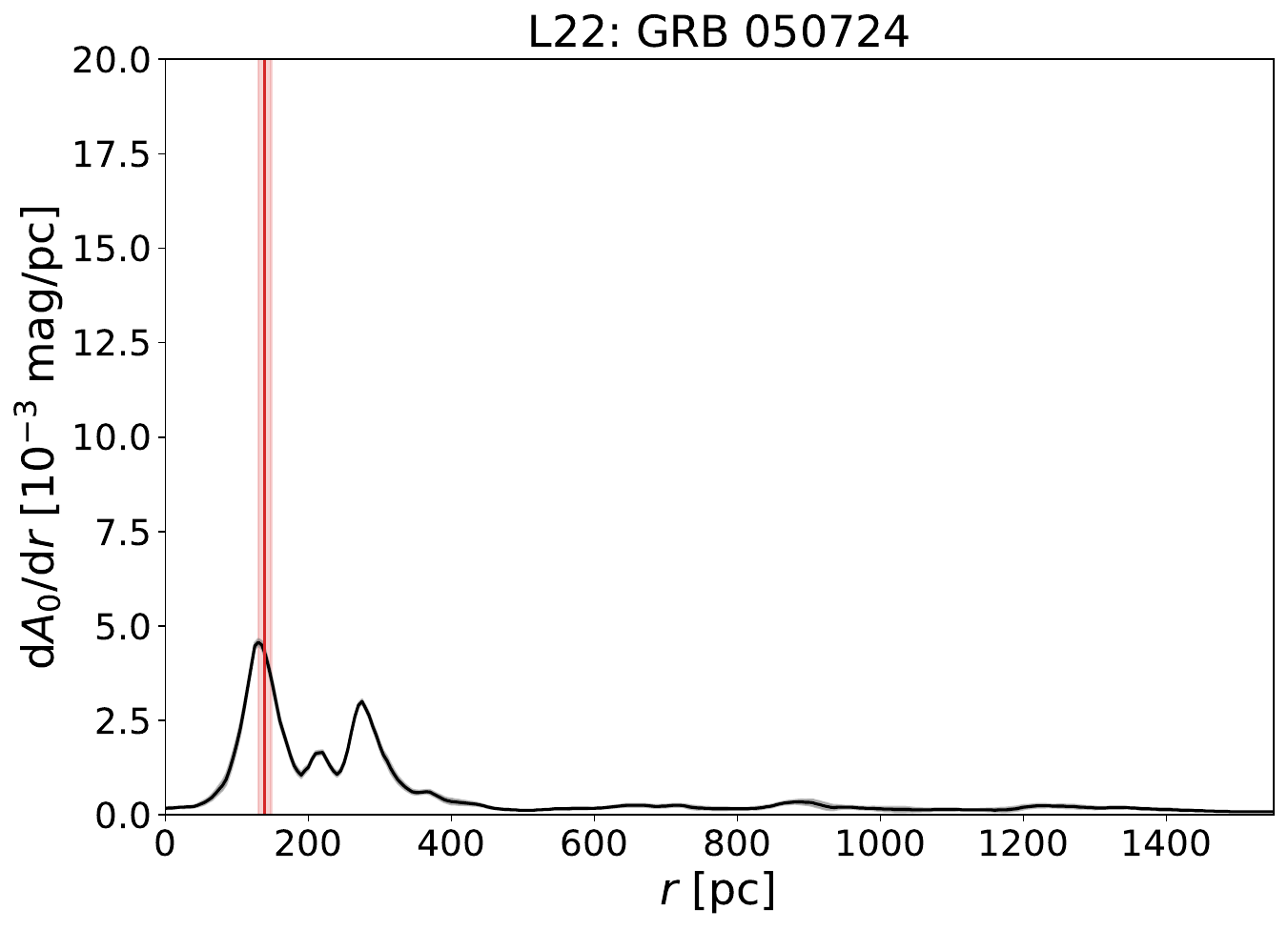}
\includegraphics[width=0.40\textwidth]{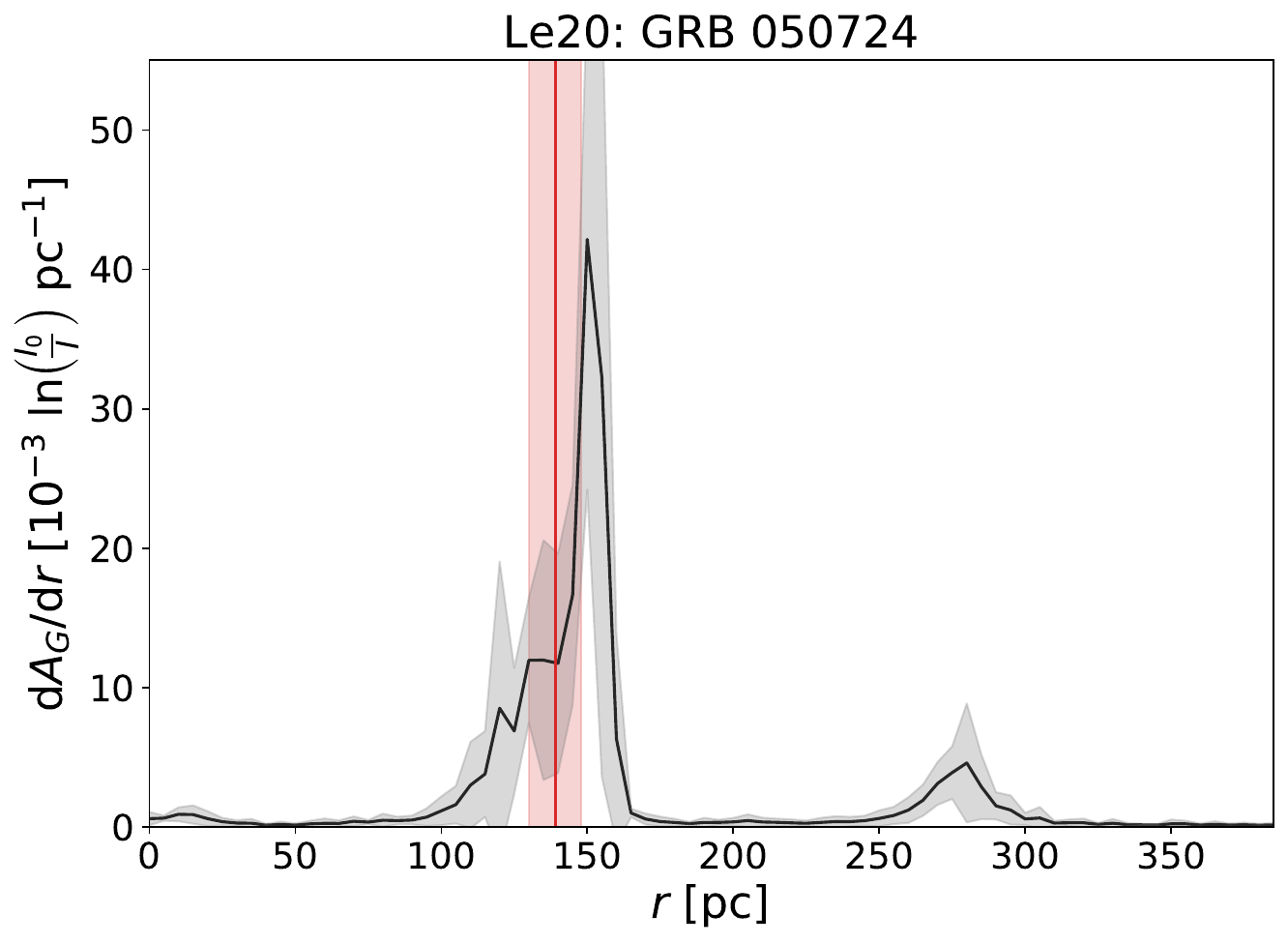}
\includegraphics[width=1\textwidth, height=4.8cm] {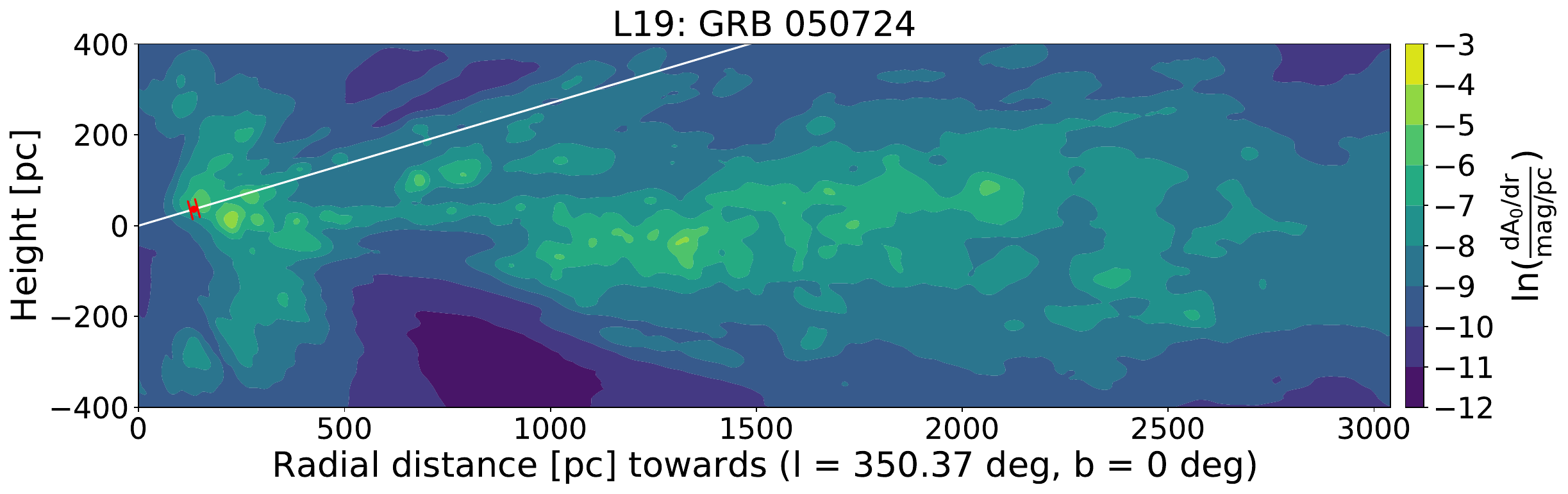}
\includegraphics[width=1\textwidth, height=4.8cm]{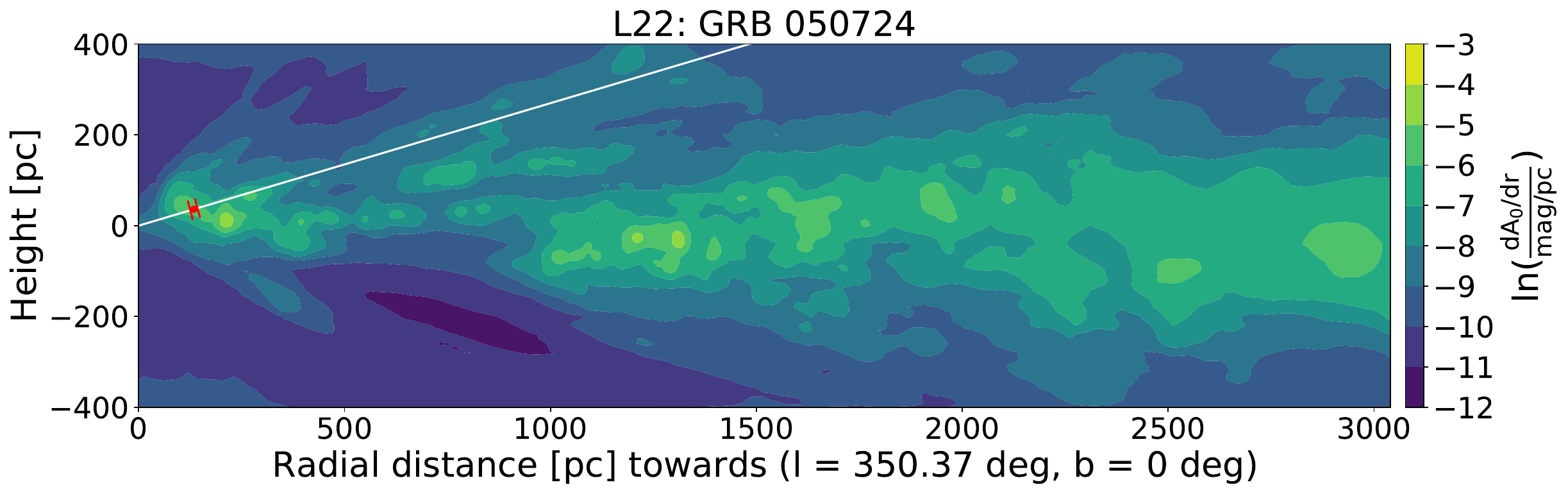}
\captionof{figure}{GRB 050724. First and second row same as in Fig. \ref{fig:combo-160623A}, without G19 map. Last two rows same as in Fig. \ref{fig:Lall2D-160623A}.}
\label{fig:appendix_050724}

\includegraphics[width=0.40\textwidth]{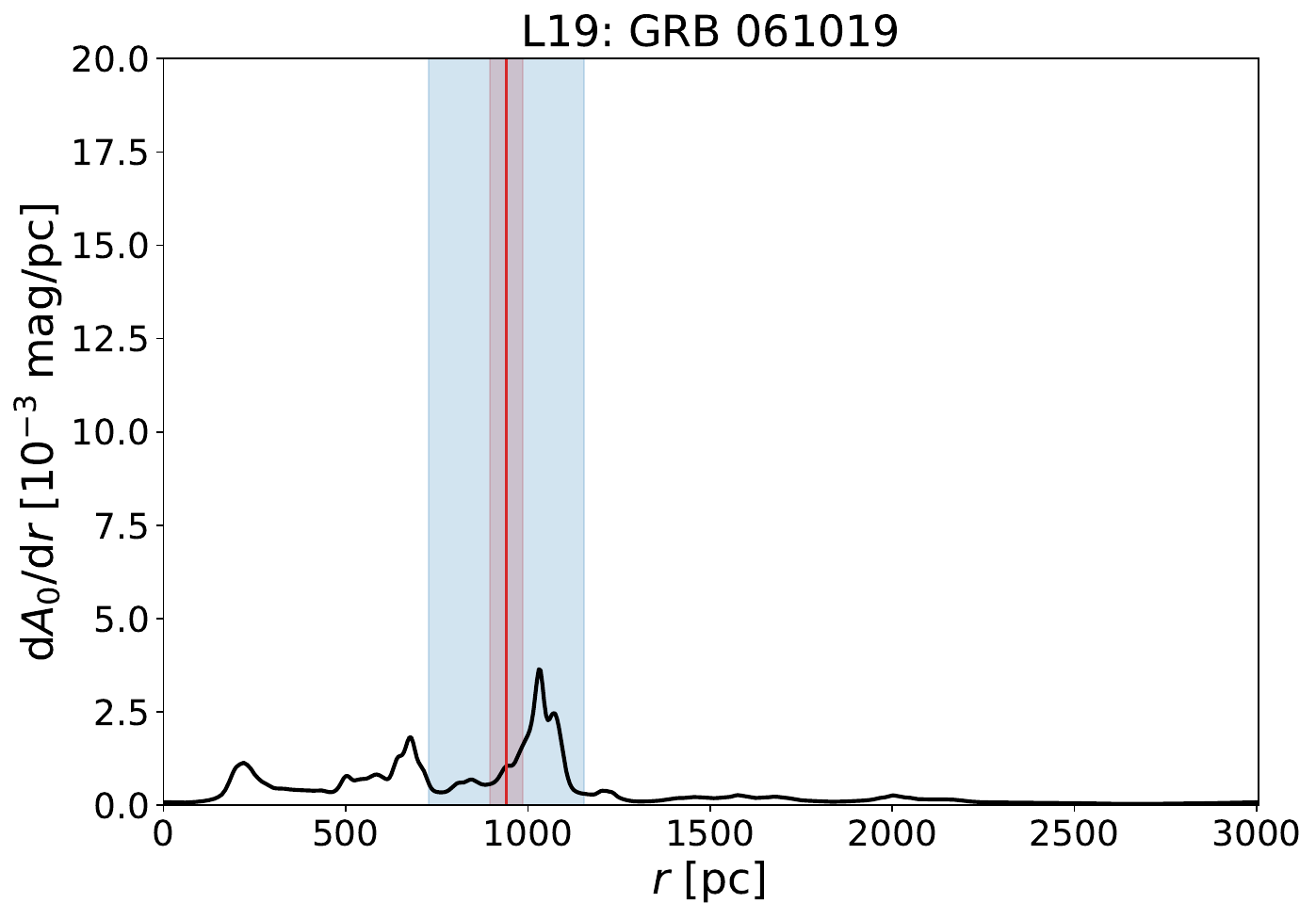}
\includegraphics[width=0.40\textwidth]{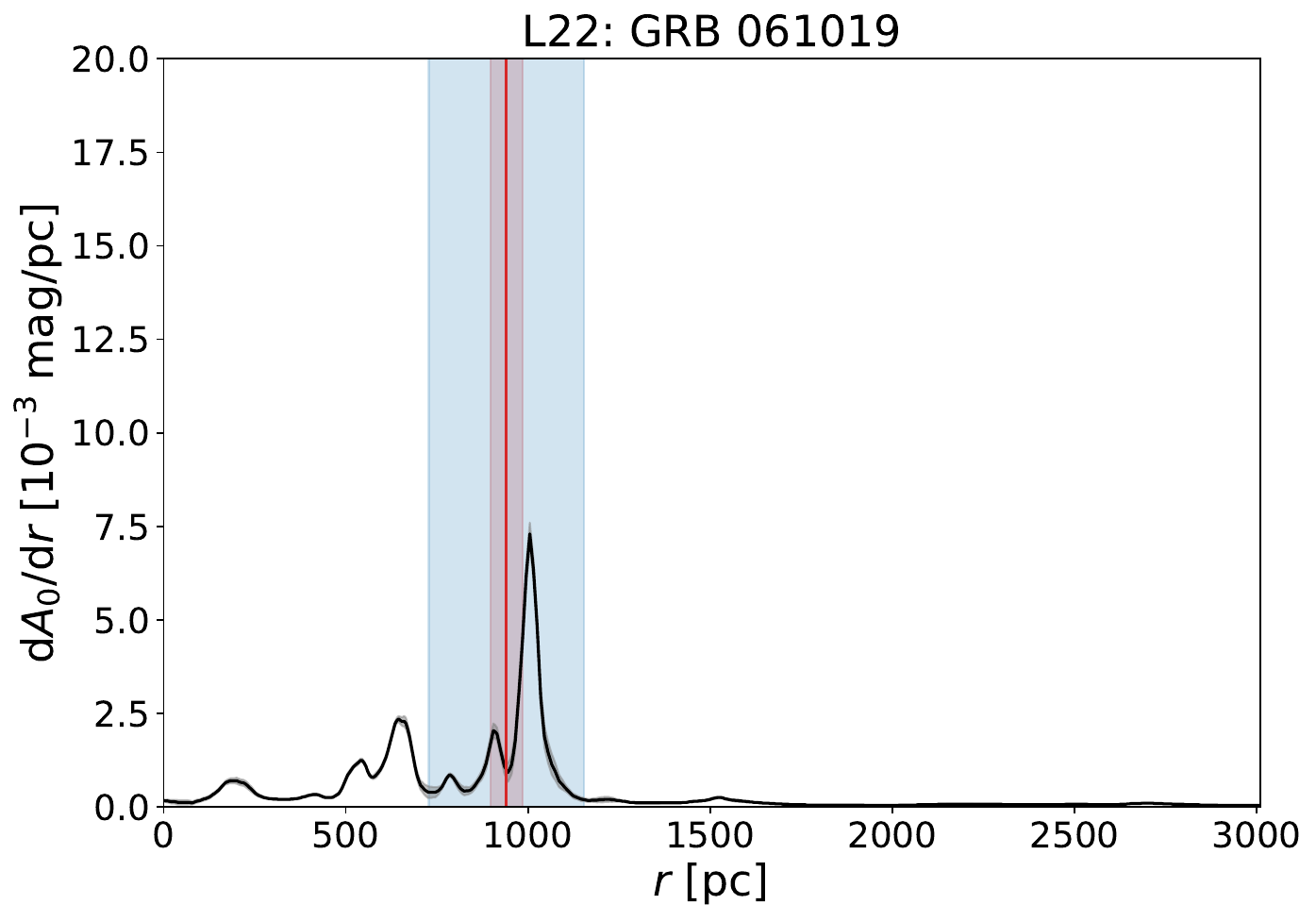}
\includegraphics[width=0.40\textwidth]{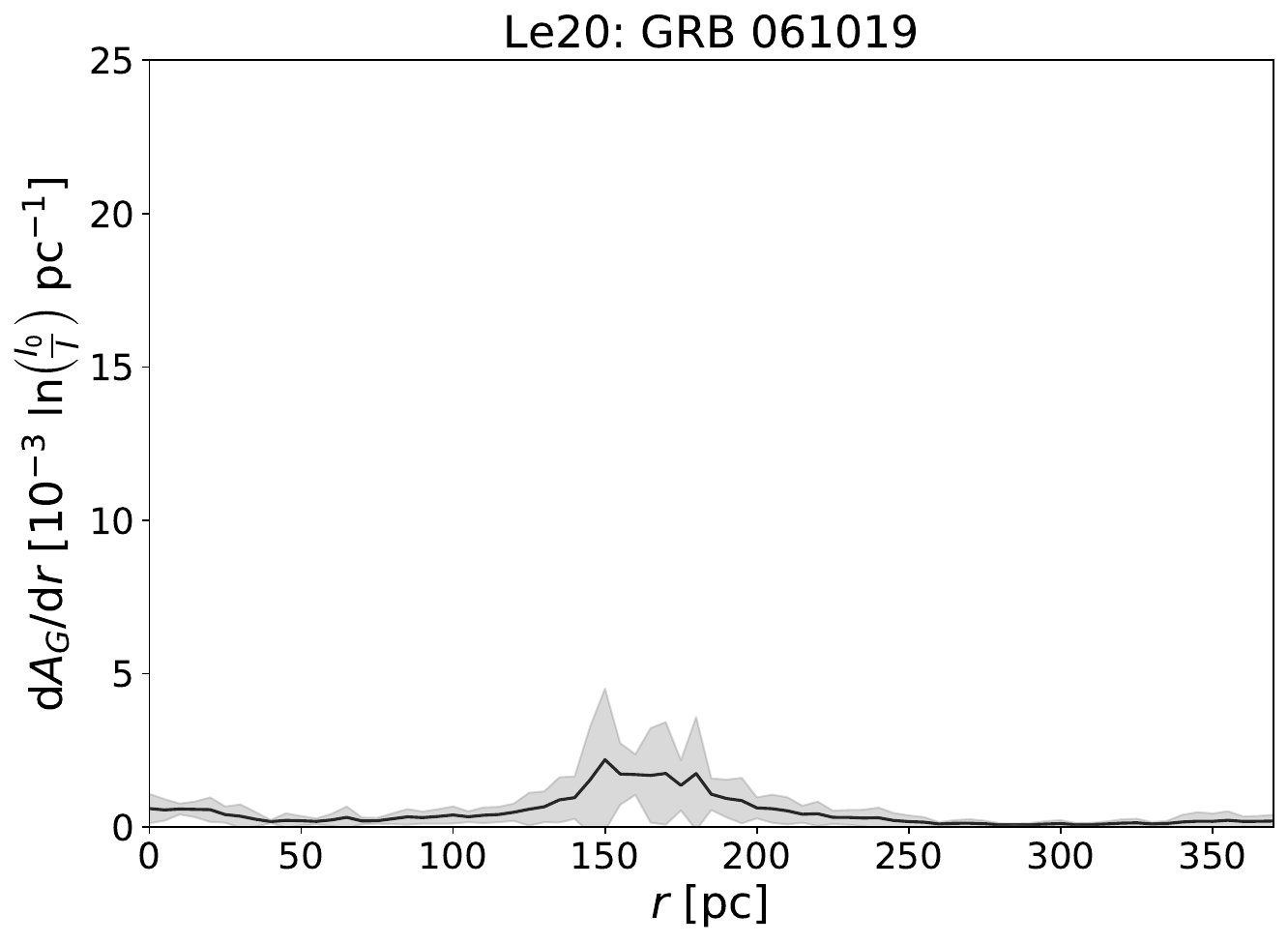}
\includegraphics[width=1\textwidth, height=4.8cm] {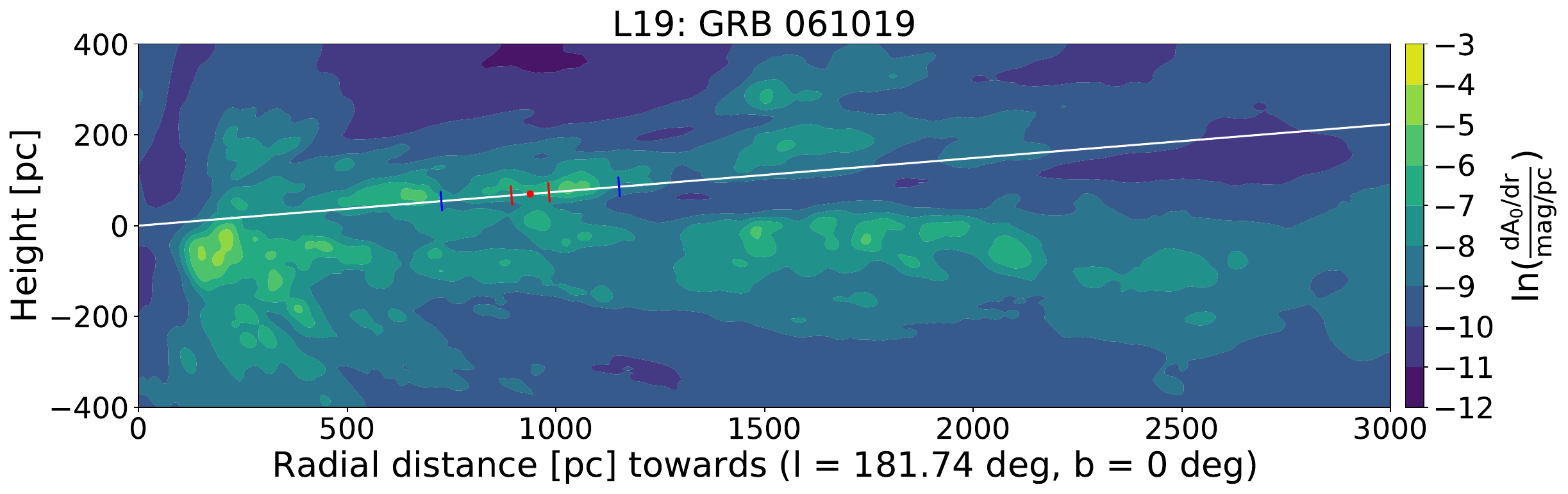}
\includegraphics[width=1\textwidth, height=4.8cm]{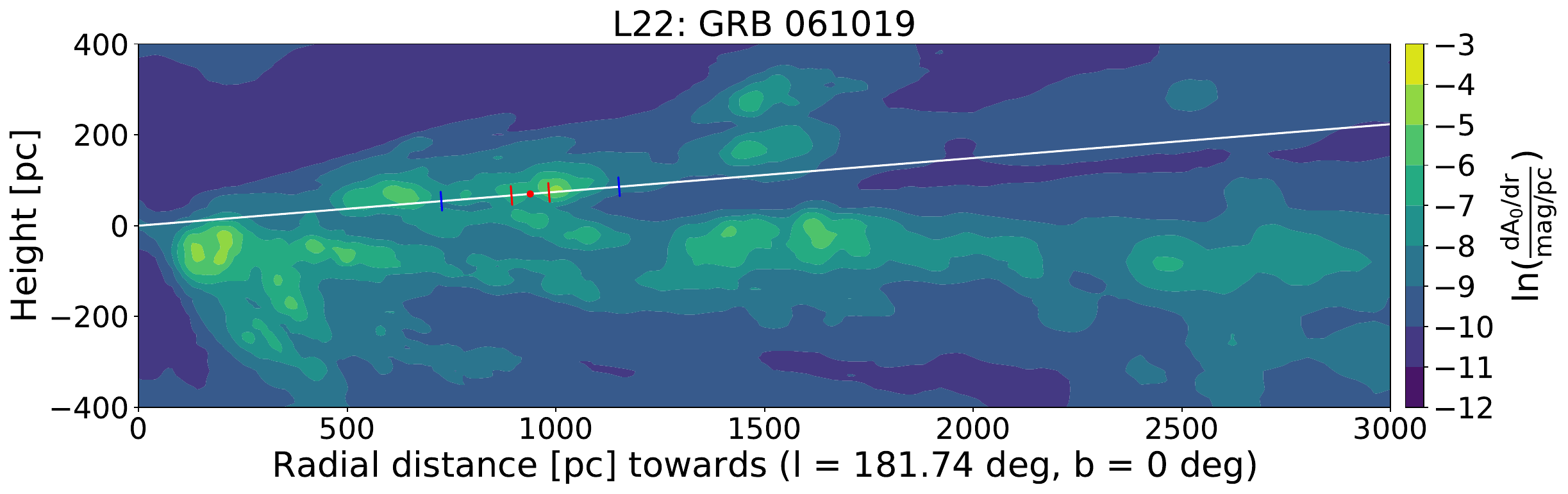}
\captionof{figure}{GRB 061019. First and second row same as in Fig. \ref{fig:combo-160623A}, without G19 map. Last two rows same as in Fig. \ref{fig:Lall2D-160623A}.}
\label{fig:appendix_061019}

\includegraphics[width=0.40\textwidth]{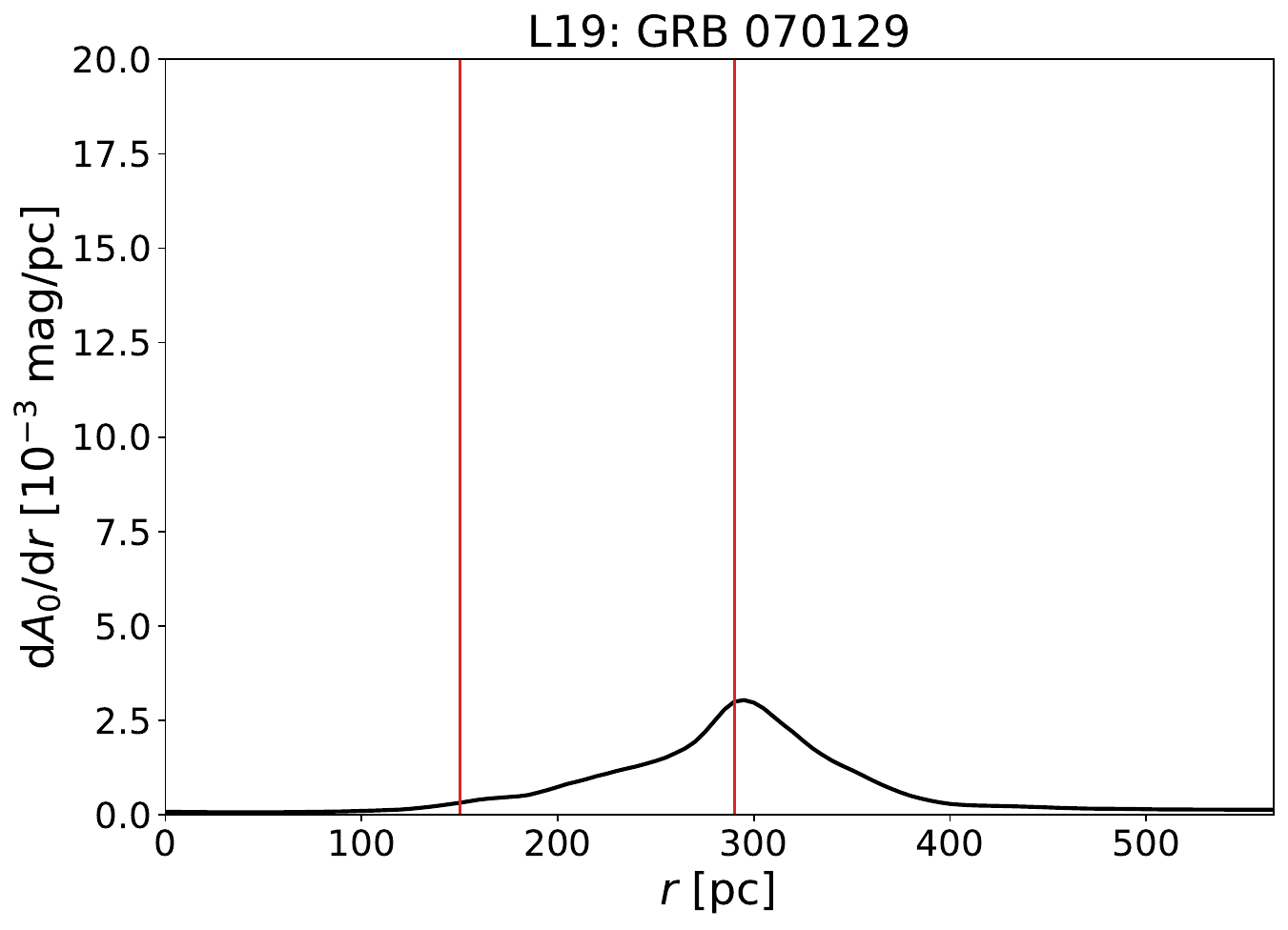}
\includegraphics[width=0.40\textwidth]{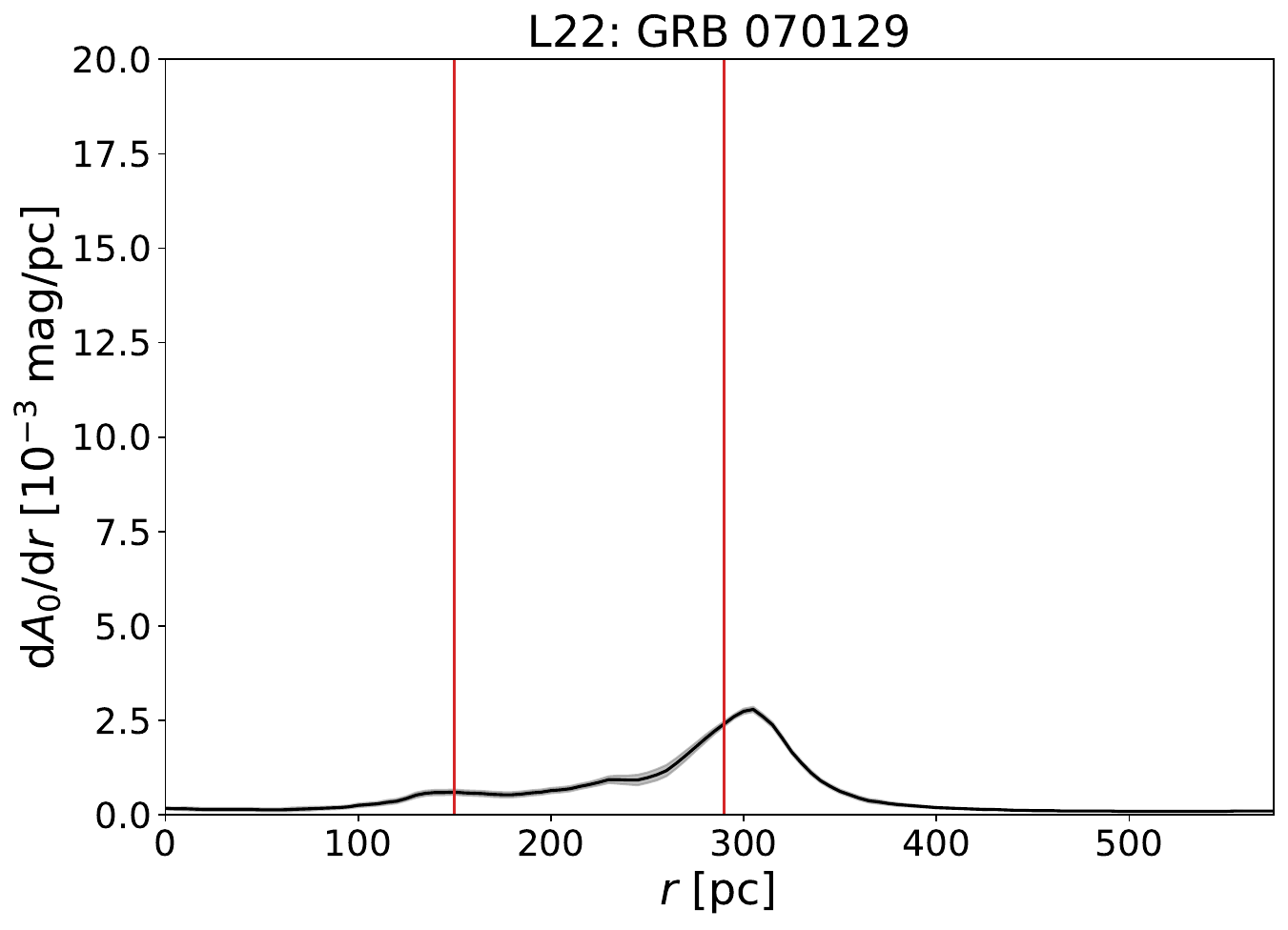}
\includegraphics[width=0.40\textwidth]{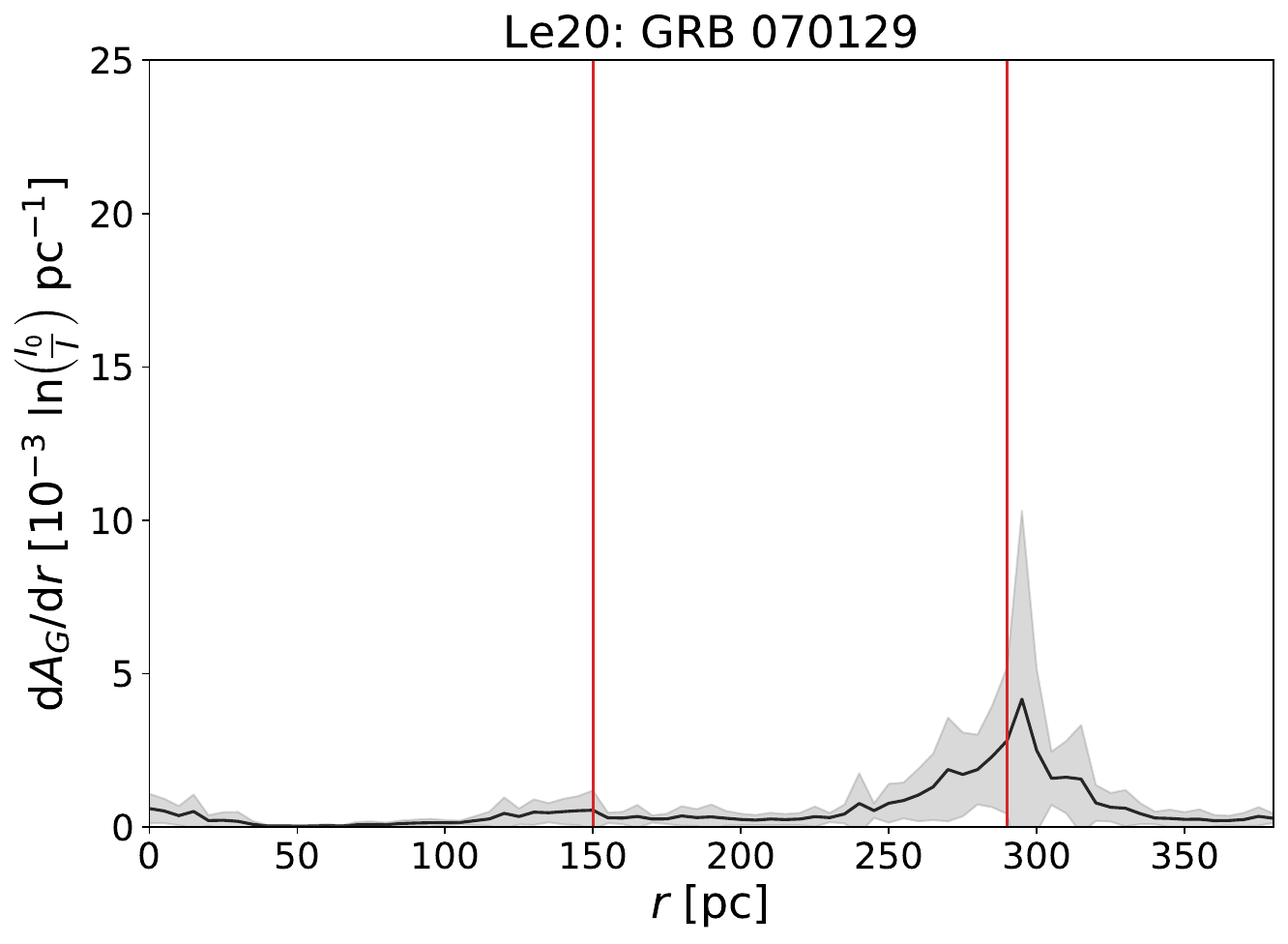}
\includegraphics[width=1\textwidth, height=4.8cm] {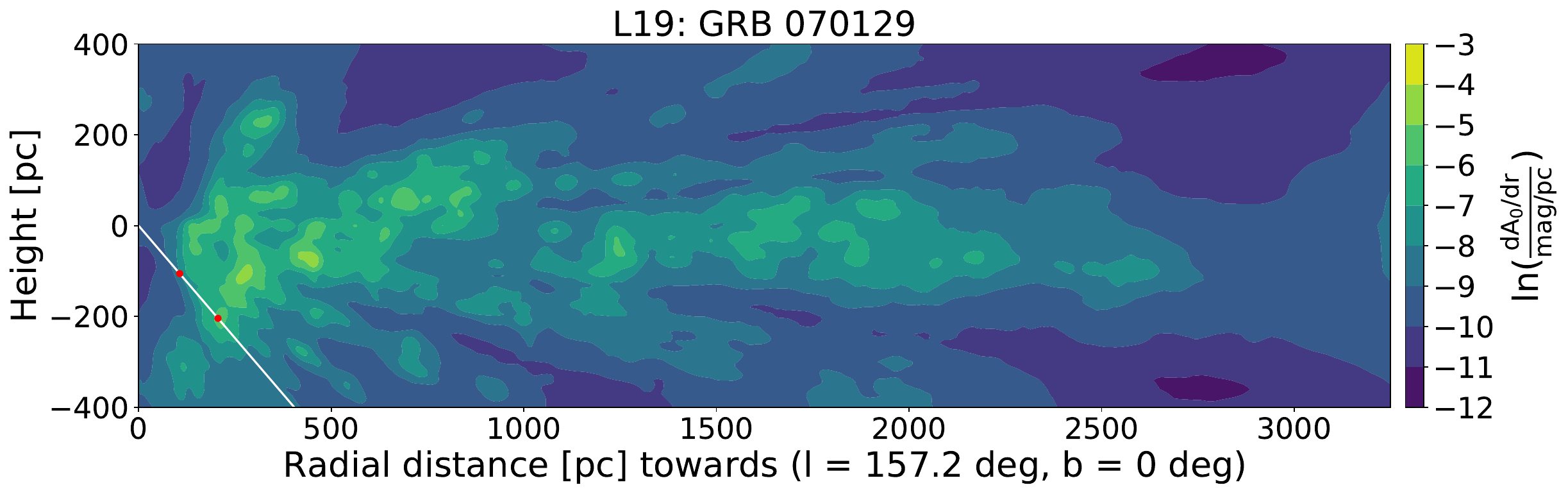}
\includegraphics[width=1\textwidth, height=4.8cm]{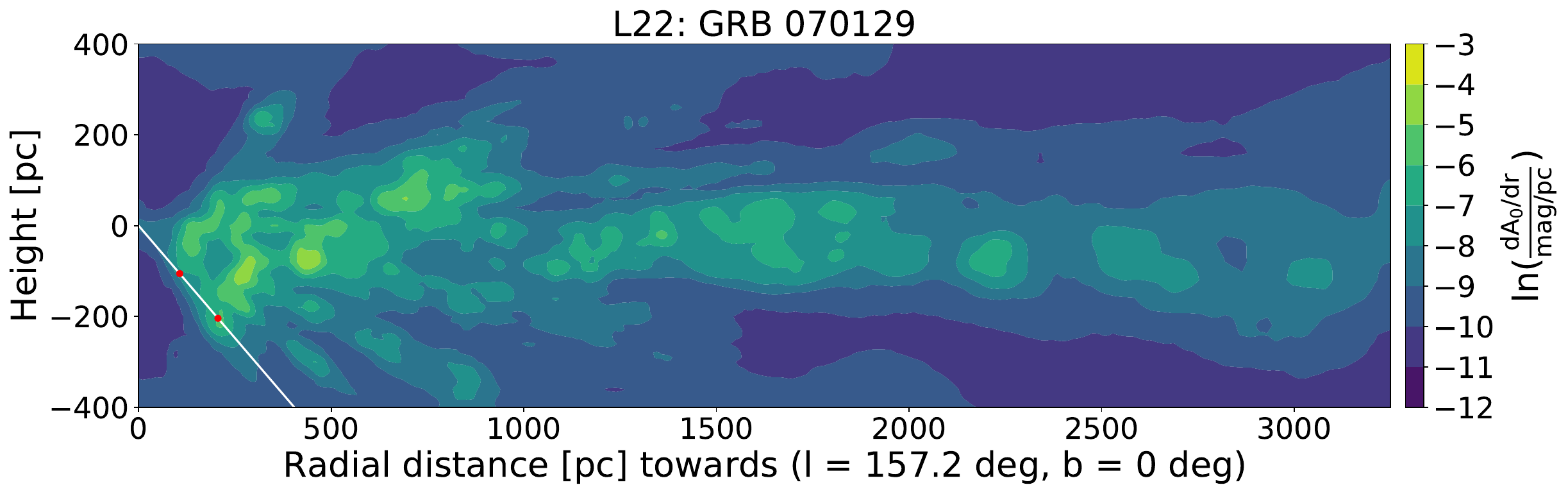}
\captionof{figure}{GRB 070129. First and second row same as in Fig. \ref{fig:combo-160623A}, without G19 map. Last two rows same as in Fig. \ref{fig:Lall2D-160623A}.}
\label{fig:appendix_070129}

\includegraphics[width=0.40\textwidth]{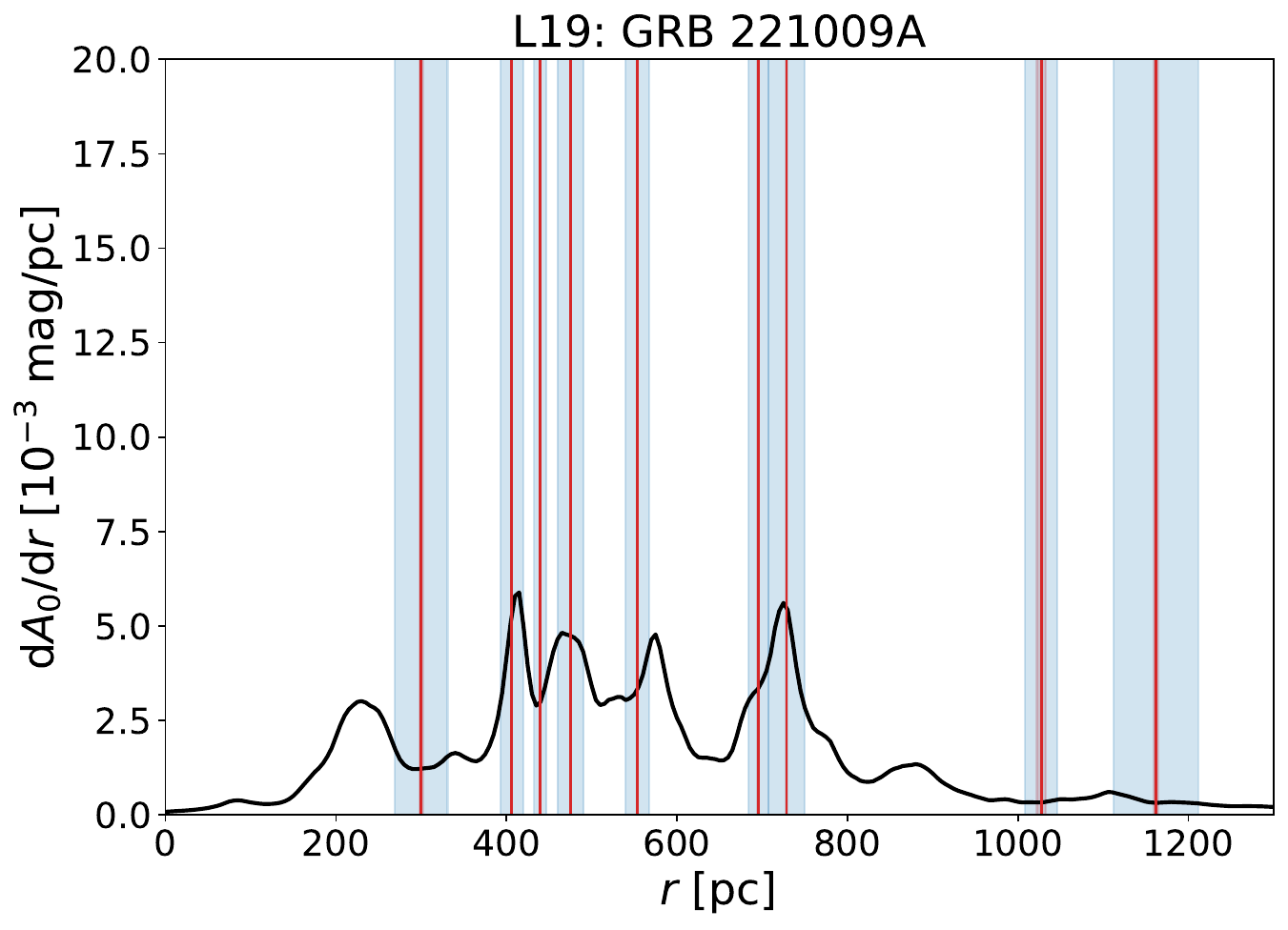}
\includegraphics[width=0.40\textwidth]{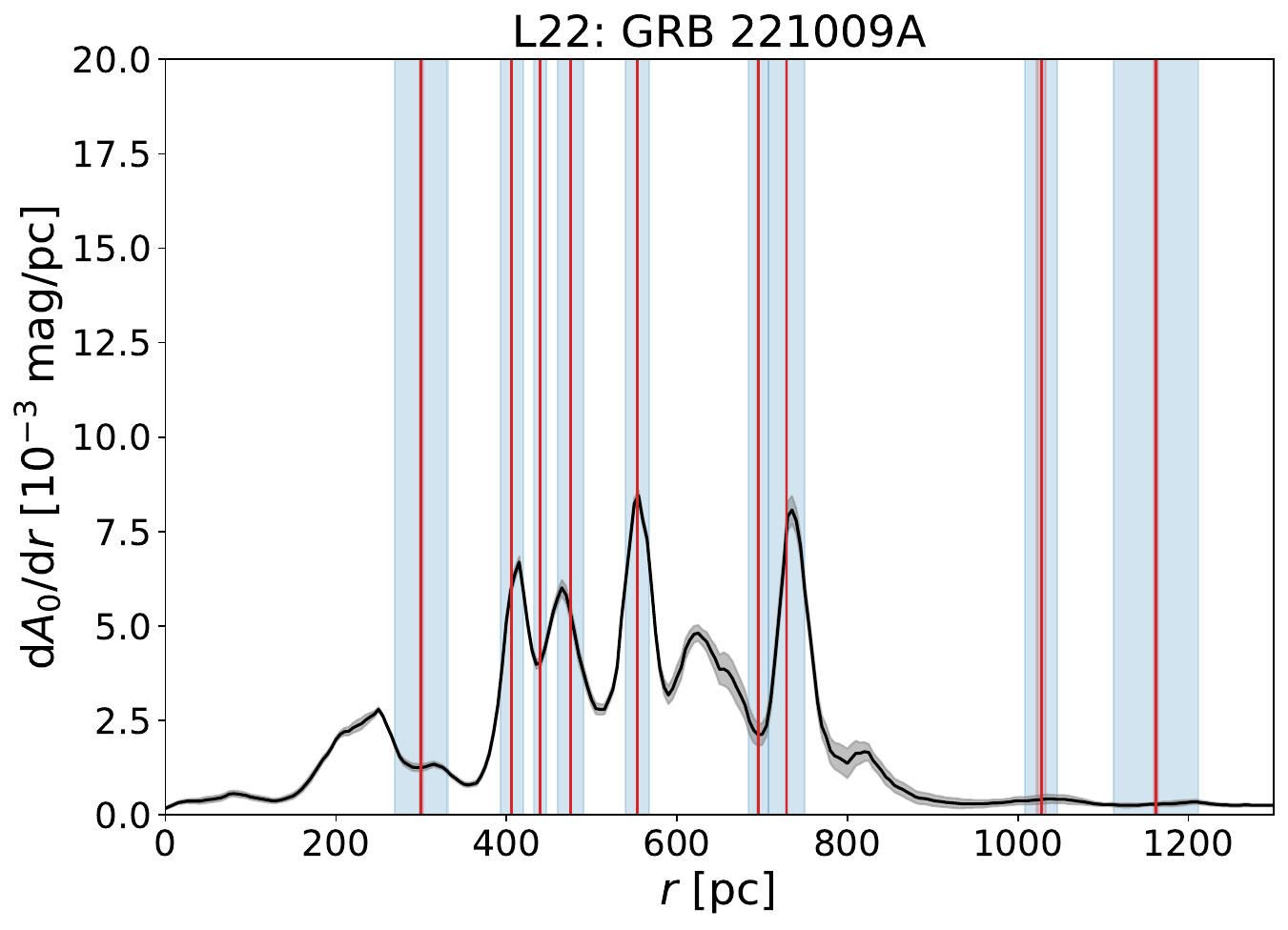}
\includegraphics[width=0.40\textwidth]{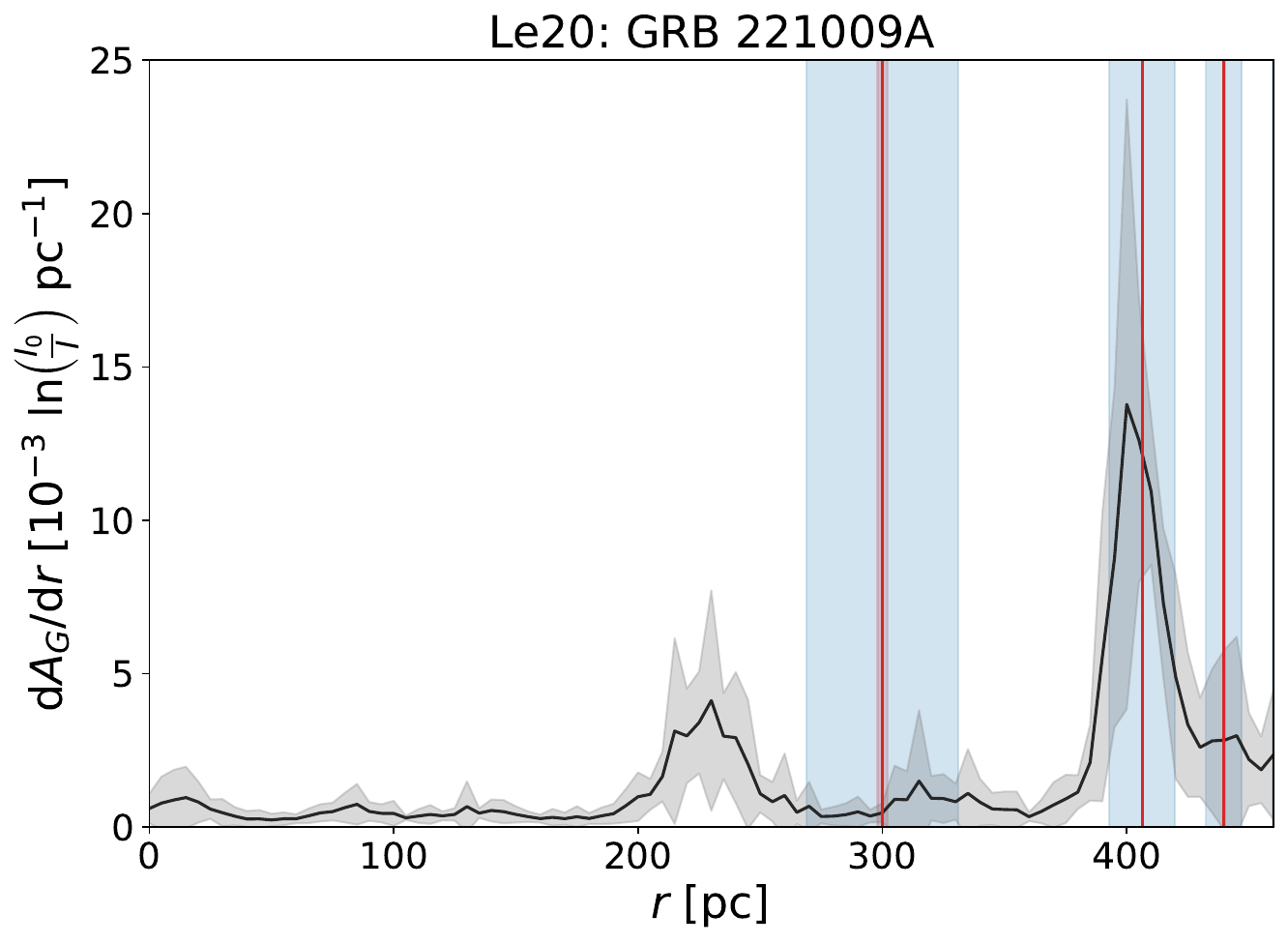}
\includegraphics[width=1\textwidth, height=4.8cm] {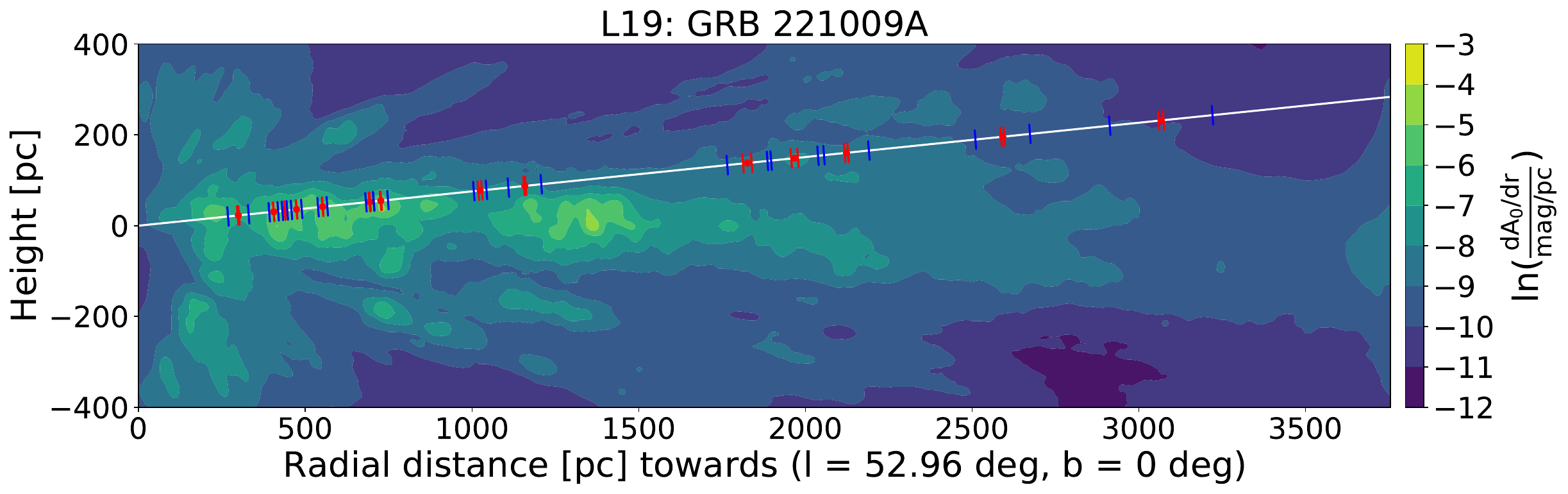}
\includegraphics[width=1\textwidth, height=4.8cm]{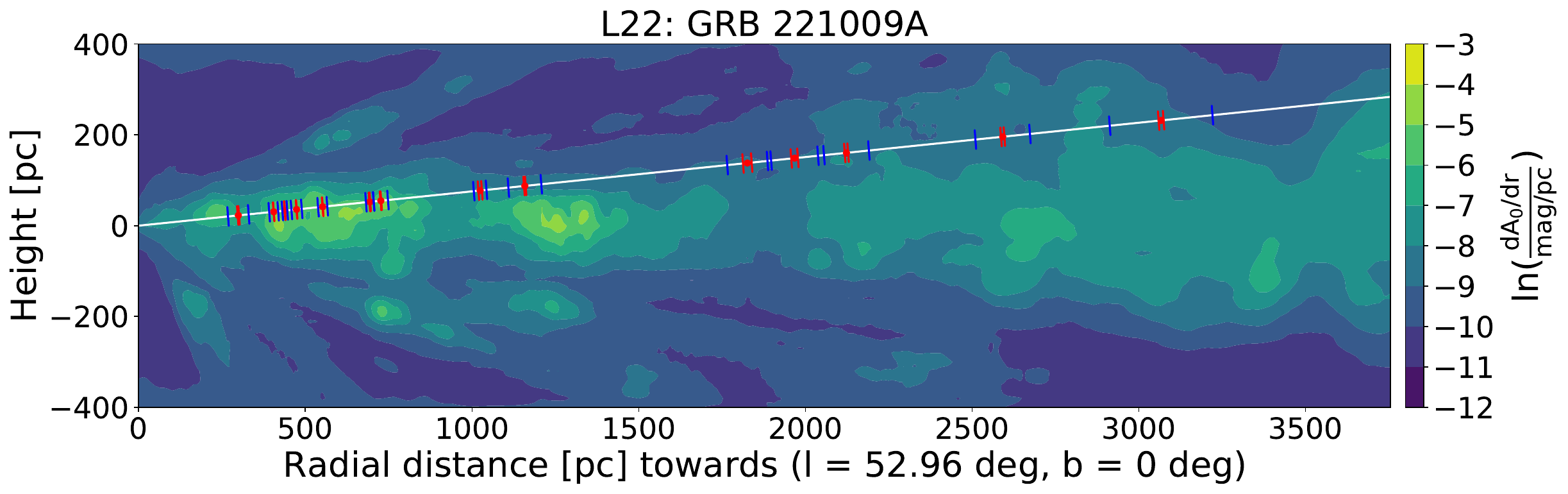}
\captionof{figure}{GRB 221009A. First and second row same as in Fig. \ref{fig:combo-160623A}, without G19 map. The L19 and L22 extinction density distributions are plotted only until 1300 pc in order to better resolve X-ray measurements at shorter distances. We note that at larger distances, there are no peaks in extinction corresponding to X-ray measured positions of dust layers, as in the case of GRB 160623A (Fig. \ref{fig:combo-160623A}). Last two rows same as in Fig. \ref{fig:Lall2D-160623A}.}
\label{fig:appendix_221009A}

%%%%%%%%%%%%%%%%%%%%%%%%%%%%%%%%%%%%%%%%%%%%%%%%%%

% Don't change these lines
\bsp	% typesetting comment
\label{lastpage}
\end{document}